\documentclass[sigconf,nonacm]{acmart}  %%
\usepackage{subcaption, algorithm, algorithmic, float, multirow, makecell}
\newcommand{\excise}[1]{}

\AtBeginDocument{%
  \providecommand\BibTeX{{%
    \normalfont B\kern-0.5em{\scshape i\kern-0.25em b}\kern-0.8em\TeX}}}

\begin{document}
%\tableofcontents
%\newpage
\title{Distribution and Management of Datacenter Load Decoupling} % (for the Grid)}
%\title{Site-aware and Grid-interactive Datacenter Load Decoupling}
%title{Placing Wisely and Playing Efficiently with Datacenter Load Decoupling}

\begin{abstract}
% compress to 1 sentence
The exploding power consumption of AI and cloud datacenters (DCs) intensifies the long-standing concerns about their carbon footprint, especially because DCs' need for constant power clashes with volatile renewable generation needed for grid decarbonization. DC flexibility (a.k.a. load adaptation) is a key to reducing DC carbon emissions by improving grid renewable absorption.
%AI and Cloud Datacenter (DC) growth is limited by power grid capacity because DC's need for constant power clashes with the variation inherent in volatile renewable generation. 
%Adapting DCs' load to grid dynamics can resolve this conflict, but how to realize the load flexibility and effectively utilize this flexibility for global benefits is challenging.
%DC flexibility (a.k.a. load adaptation) is a key strategy that can quickly increase grid capacity for datacenters. 

DC flexibility can be created, without disturbing datacenter capacity by \textit{decoupling} a datacenter's power capacity and grid load with a collection of energy resources.  Because decoupling can be costly, we study how to best distribute and manage decoupling to maximize benefits for all.  Key considerations include site variation and datacenter-grid cooperation.

% raise questions of "how to distribute", "how to manage efficiently"
% we study this, that
% and results

%% topically correct, but the wrong way to talk about it
%We formulate the problem of DC grid load adaptation with decoupling, covering the whole process from resource planning to benefit allocation.

We first define and compute the power and energy needs of datacenter load decoupling, and then we evaluate designed distribution and management approaches. Evaluation shows that optimized distribution can deliver >98\% of the potential grid carbon reduction with 70\% of the total decoupling need. % 70\% of the total decoupling need vs. unconstrained cases. 
For management, DC-grid cooperation (2-way sharing and control vs. 1-way info sharing) enables 1.4x grid carbon reduction. Finally, we show that decoupling may be economically viable, as on average datacenters can get power cost and carbon emissions benefits greater than their local costs of decoupling.  However, skew across sites suggests grid intervention may be required.

%monetize the benefits and costs for the grid as a whole and individual datacenters. The designed approaches can achieve benefits higher than costs for both the grid and DCs, incentivizing the realization of decoupling.
\end{abstract}

\author{Liuzixuan Lin}
\affiliation{
  \institution{University of Chicago}
  \streetaddress{5730 S Ellis Ave}
  \city{Chicago}
  \state{IL}
  \country{USA}
  \postcode{60637}
}
\email{lzixuan@uchicago.edu}

\author{Andrew A. Chien}
\affiliation{
  \institution{University of Chicago \& Argonne National Lab}
  \streetaddress{5730 S Ellis Ave}
  \city{Chicago}
  \state{IL}
  \country{USA}
  \postcode{60637}
}
\email{aachien@uchicago.edu}








%\tableofcontents
%\settopmatter{printfolios=true}
\maketitle

\section{Introduction}
\label{sec:intro}
Datacenters are one of the fastest-growing sectors of power consumption in many regions around the world \cite{iea2024electricity}. In the U.S., from 2018 to 2022, datacenters' power consumption grew 17\% annually, driven by rapid cloud growth.  However, the annual growth has doubled since 2022, with projected growth of 27\%/year through 2028 and beyond \cite{lbnl2024DCUsage}.  Similar growth is projected  worldwide \cite{mckinseyExpandingDC}. This growth is fueled by expanding AI applications and cloud business, reflected in hyperscalers' exploding capital expenditures (CapEx). The total CapEx of the 11 largest hyperscalers in the world has grown to \$392B in 2025.  This number is a record high that exceeds the annual investments in 2022 and 2023 combined \cite{techCapexGrowth2025May}. 
%In Europe and North America, this rapid DC growth has begun to be limited by the ability to connect to power grids for reliable power\cite{epriUtilitySurvey2024,eirgrid23report,DominionBraceForAI23}.

This accelerated growth intensifies the long-standing concerns about datacenters' growing power consumption and associated carbon footprint \cite{Masanet20,Greenpeace-Nova19}. In addition to the growth, a root cause of the concerns is the conflict between datacenters' need for constant power and the variation inherent in volatile renewable generation needed for grid decarbonization. To avoid decarbonization slow-down and to protect grid reliability, some grid operators delay approval or even suspend new datacenter projects \cite{SingaporeCondLiftDCPause,IrelandHaltDC}. In some other cases, the grid delay the decommission of coal power plants or propose building new gas power plants to support datacenter growth \cite{midwestDCEnergyChallenge,AI-disrupt-green-dc-efforts,goldmanSachs2024}, which provide reliable power supply at low cost but clearly destructive to grid decarbonization. Overall, this conflict put either datacenter growth or grid decarbonization (and greening compute along with it) at risk.

\paragraph{\textbf{Need for Datacenter Flexibility}} Hyperscalers have sustainability commitments such as ``net-zero by 2030'', which require a significant decrease in datacenter carbon emissions. Although they are purchasing renewable power through power purchase agreement (PPA) to ``offset'' their power consumption, the increased need of new  gas generators in the grids and the 15--20\% annual growth of Scope 2 carbon emissions reported \cite{googleEnv2025,microsoftEnv2025} reflect datacenters' heavy reliance on fossil-fuel generation. Datacenter flexibility to match load to renewable generation is the key to achieving DC decarbonization goals.  In power grids today where many include 10--40\% renewables with growth to 40--90\% by 2030 \cite{iea2024renewables}, significant renewable generation is wasted (``curtailed'') or sold at zero or negative prices \cite{chien2024grids,ERCOT-stranded20,ye2018wind,Germany-negative-price16}. For example, curtailment in CAISO (California) was 3.4 TWh in 2024, having grown 40\%/year in 2015--2023 \cite{chien2024grids}.  Further, gas generation powers datacenters at night, or when wind generation is insufficient. Datacenter flexibility can reduce both the curtailment and the use of fossil-fuel generation.
%Several recent studies show that even rare load shedding (involuntary reduction) can increase grid datacenter-capacity \cite{lin2024exploding,norris2025rethinking}.  The magnitude of this effect is large; load flexibility approaches could enable U.S. grids to accommodate 5-10 years of datacenter growth without new generation or transmission \cite{norris2025rethinking}.  Because of this reality, grids have begun to require load flexibility as a condition for DC grid connection \cite{eirgrid23report,pgeFlexConnect,pjmFlexLoadProposal}. For example, PJM, which includes ``Datacenter Alley'' in Northern Virginia, recently proposed a new program that allows datacenters willing to accept reduced power reliability not to purchase long-term capacity for grid connection \cite{pjmFlexLoadProposal}. Alberta has mandated flexibility \cite{alberta}, and  voluntary shedding agreements have been undertaken in Indiana and Minnesota \cite{google-article-tn-in}.
Finally, the promise of datacenter flexibility to reduce DC carbon emissions has been shown by extensive research  \cite{lin2023adapting,luo2013temporal,liu2012renewable,dou2017carbon,goiri2013parasol,hanafy2023carbonscaler,deng2014harnessing,sukprasert2024limitations,souza2023ecovisor}. However, because computing workloads traditionally require constant power, the adoption of datacenter flexibility in industrial practice is rare or on a limited scale. 

\paragraph{\textbf{Need for Decoupling}} Cloud datacenters' need for constant power is inherent in quality of service (QoS) commitments (e.g. 99.999\% availability) and business pressure for maximum utilization. Varying datacenter power capacity to match volatile renewable generation harms datacenter productivity \cite{xing2023carbon,zhang2021scheduling}, which perhaps explains the limited flexibility in Google's Carbon-aware Computing \cite{radovanovic2021carbon}. Furthermore, traditional load flexibility incentives including time-varying power rate and incentive payments are not attractive to cloud datacenters as power cost is only a small fraction relative to the revenue of cloud services \cite{Barroso18,cloudMargin}.  

\textbf{What if we could achieve both stable datacenter capacity and high renewable power absorption?}  That would maintain high IT efficiency while enabling the power grid to decarbonize.
For backup, %Facing the requirement of load flexibility, 
many datacenters deploy energy resources such as generators and batteries \cite{broadInterestInNuc,msftIEGasGenforDC,xAIMegapack25}.  Such energy resources can be used not just in emergencies, but more generally to \textit{decouple} datacenter's power capacity from its grid load.  This enables IT operation to be unaffected while the grid load is flexed to avoid carbon emissions.
\textbf{But how much energy resource should be deployed and where?}
With complex grid structure and dynamics, deciding whether additional energy resources are beneficial locally is difficult.  It's even more challenging to understand if they would be more valuable at another datacenter site. 
\textbf{Beyond distribution, intelligent management is another challenge.} As large loads up to gigawatts \cite{MetaBuildingGWDC} and collectively accounting for 20--30\% of grid load \cite{dominionVA23report,eirgrid23report}, datacenters that flex grid load to match renewable generation in the grid can overshift their load, disrupting grid dispatch \cite{lin2023adapting}.
%The better measurability of energy resource cost for decoupling vs. IT workload impact enables a more quantitative comparison of load adaptation's costs and benefits. 
Such management would require intelligent coordination between DCs and the grid to achieve maximum benefits.
%approaches to achieve low costs and high benefits from load adaptation.

\begin{figure}[h]
    \centering
    \includegraphics[width=\columnwidth]{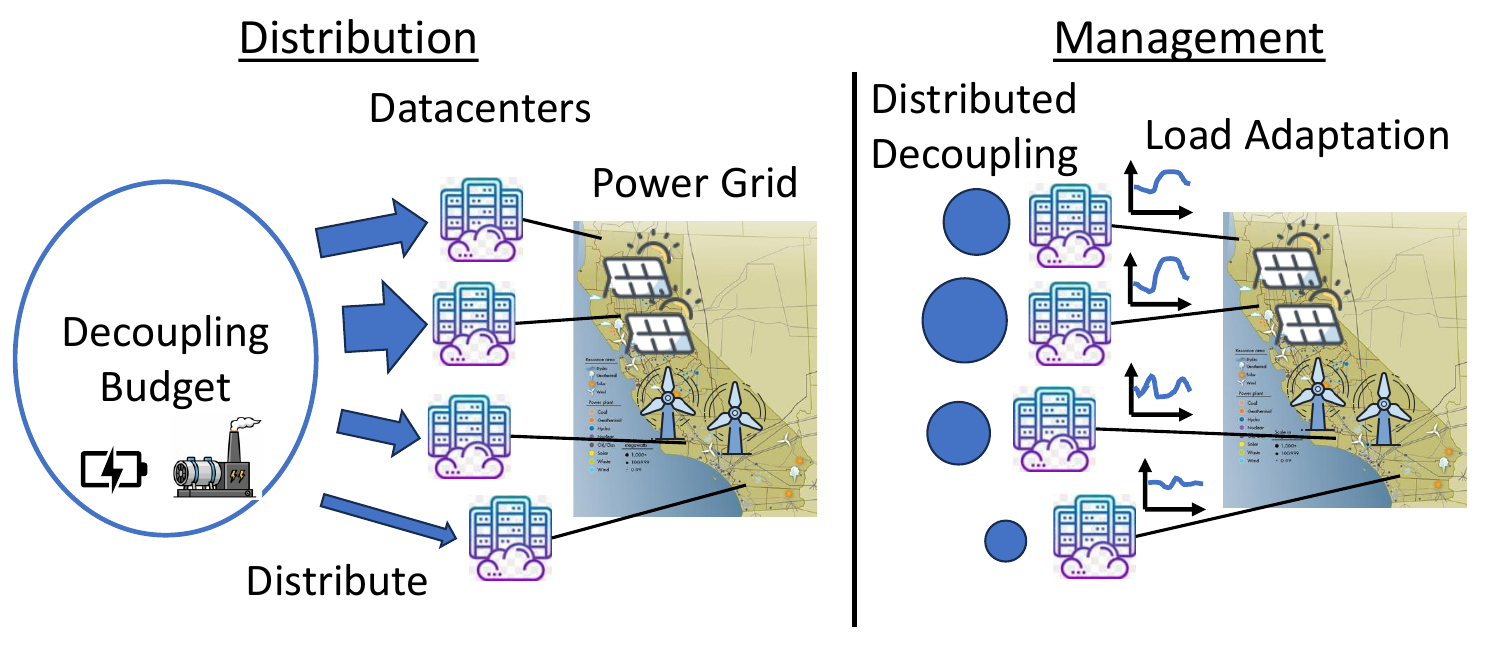}
    \caption{Two Phases of Datacenter Load Decoupling: (1) provision energy resources for decoupling at various datacenter sites; (2) manage decoupling to adapt datacenter grid load to grid dynamics.}
    \label{fig:intro}
\end{figure}

With the goal of maximizing DC and grid carbon reduction at low cost, we study the distribution and management of decoupling energy resources.  We first analyze grid dynamics and ideal load profiles to understand decoupling requirements.  Then, as shown in Figure \ref{fig:intro}, we distribute a total decoupling budget across datacenters for maximum potential benefits.  Finally, we consider coordination and management approaches that exploit distributed decoupling efficiently for carbon emissions reduction. Specific contributions include:
\begin{itemize}
    \item Definition and computation of datacenter decoupling power and energy needs.
    \item Given a global decoupling budget, define a method that distributes it for high effectiveness.  Specifically, our stochastic optimization approach delivers 98--100\% of potential grid carbon reduction with only 70\% of the total decoupling need. % vs. unconstrained cases.
    \item Design and evaluation of decoupling management approaches. We propose PS-GridScale, a new 2-way sharing and control approach, comparing it with 1-way information sharing (PlanShare) and grid-optimal control approaches. PS-GridScale achieves up to 1.4x further grid carbon reduction vs. PlanShare and captures 84--90\% of the potential benefits. % with grid-optimal control.
    \item Financial analysis that shows decoupling may be economically viable.  On average, DCs gain power cost and carbon emissions benefits greater than their local costs of decoupling.  However, the skew across sites suggests some grid intervention may be required for implementation.  In addition, we show power grids may gain more decarbonization from adding decoupling than building additional renewable generation.
    
%    Cost-Benefit analysis for the grid as a whole and individual datacenters, which sheds light on who has incentives to implement decoupling. We monetize both costs and benefits, showing that our overall approach achieves benefit higher than decoupling cost for both the grid and DCs. 
\end{itemize}

The remainder of the paper is organized as follows. Section \ref{sec:background} introduces the background of need for datacenter flexibility and decoupling. Section \ref{sec:problemAndApproach} motivates the study of distribution and management of decoupling and summarizes our approach. And then we elaborate the design of distribution and management approaches in Section \ref{sec:loadAdaptation}, followed by corresponding evaluation in Section \ref{sec:evaluation}. Finally, we discuss the related work and conclude.

\section{Background}
\label{sec:background}
\subsection{Power Grid Decarbonization}
In response to global warming, power grid decarbonization, which replaces fossil-fuel generation with carbon-free energy, is taking place all over the world \cite{California25MMTCarbonBy2035,NY70by30,EUClimateNeutralBy2050,UKNetZeroRoadMap2021,china2023carbonPolicies}. For example, aiming at 80\% renewables in electricity by 2030, Germany reached a record share of 62.7\% in 2024 \cite{germany2024renewables}. China, the world's largest electricity producer, built 357 GW of solar and wind generation capacity in 2024 \cite{china2024renewableBuildout}. After seeing 100\% load supported by renewables for the first time in 2022, California now regularly sees this phenomenon with annual average renewable fraction of 39\% \cite{caiso2024MonthlyRenewable}. Globally, IEA projects that renewable capacity is expected to increase over 5,520 GW during 2024--2030, 2.6 times more than the deployment of the previous six years (2017--2023) \cite{iea2024renewables}.

The growth of renewable generation capacity is dominated by wind and solar that depend on natural dynamics of wind and sun. As the fraction of renewable generation increases, the challenge of balancing demand and generation increases: when renewable generation is low, power outage can happen; when renewable generation is high, it may be wasted (so-called ``curtailment'') \cite{AIMS18,ERCOT-stranded20,chien2024grids,ye2018wind} or produce negative prices in the wholesale power market \cite{Germany-negative-price16,ERCOT-negative-price15}. Supply-side solutions to balancing include energy storage \cite{caisoEnergyStorage2023}, generation capacity overprovisioning \cite{MN-solar-curtailment18,gupta2023optimal}, and peaking power plants. However, relying solely on these solutions is costly because of the long-tail statistics of wind and solar.  Demand-side solutions that adapt load to grid dynamics can significantly reduce infrastructure cost \cite{ShedShiftLBNL24}.

\subsection{Datacenters as Growing Fixed Loads}
\paragraph{Growth} Datacenters are growing in both size and share of grid load. Campuses of hundreds of megawatts are common today, and giants of multiple gigawatts are on the way \cite{MetaBuildingGWDC}. Growth as a fraction of grid load is projected to continue at current hotspots, including Northern Virginia (25\% in 2023, 49\% in 2034 projected) and Ireland (24\% in 2024, 30\% in 2032 projected).  High datacenter penetration is spreading. For example, Texas and Arizona expect 16\% and 17\% grid load from datacenters in 2030 respectively \cite{ercotLoadForecast,epriDCLoadForecast2024}. With this scale and fraction, it's important for the grids to consider datacenters when making policies and planning resources.
%Datacenters' power consumption was already growing fast (e.g. 13\% annual growth in 2018--2022 for the U.S. \cite{lbnl2024DCUsage}) before 2022 due to digitalization and cloud growth. The boom of generative AI applications since 2022 and associated huge demand for compute power has accelerated datacenter growth. In the U.S., the growth rate could be 27\% until 2028 \cite{lbnl2024DCUsage}. Globally, datacenter capacity is expected to grow 27\% annually for 2023--2028 \cite{mckinseyExpandingDC}. these growth projections are evidenced by the growing capital expenditures (CapEx) of hyperscalers, which supports the datacenter construction in the next few years. The total CapEx of the 11 largest hyperscalers in the world is forecasted to be \$392B in 2025, a record high that exceeds 2022 and 2023 combined \cite{techCapexGrowth2025May}. 

\paragraph{Fixed Grid Load} Datacenters are expensive capital assets for which hyperscalers seek to maximize utilization \cite{BorgTNG20}. The corresponding job scheduling practice (e.g. oversubscription \cite{reidys2025coach}) and service models (e.g. constant-rate VMs) don't tolerate power variation, as it can cause job termination or increased service latency  \cite{zhang2021scheduling,xing2023carbon}.  And despite a decade of effort on energy-proportionality, inflexible idle server power still corresponds to 60\% of energy consumption in commercial clouds \cite{schneider2024carbon}. Together, these result in cloud datacenters' highly fixed and inflexible grid load.%The growth of datacenters as constant loads increases the grid peak load, requiring grid infrastructure build-out. Furthermore, it's hard to be balanced with time-varying renewable generation, retarding grid decarbonization. These result in delay or even suspension of datacenter growth in North America and Europe \cite{epriUtilitySurvey2024,IrelandHaltDC,DominionBraceForAI23}.

\subsection{Realizing Datacenter Flexibility} 
%Solving this balancing challenge requires adapting datacenters' power consumption to grid dynamics. Some grid operators have propose that if datacenters can adjust grid load per grid instructions, more datacenters can be connected to the grid \cite{pgeFlexConnect,pjmLargeLoadArrangement,eirgrid23report}. In other words, the capability of grid load adaptation is now a key to the hyperscalers' business growth. 
Datacenter grid load flexibility is the key to reconciling grid decarbonization and datacenter growth as reflected in a major EPRI DCFlex initiative involving all of the leading hyperscalers and grid companies in the U.S. \cite{epriDCFlex}.  Generally, grid load flexibility can be realized by power use behavior changes and behind-the-meter energy storage \cite{ShedShiftLBNL24}, which correspond to IT operation and decoupling for datacenters. Many academic researchers have explored flexing datacenter grid load through dynamic resource scaling (e.g. toggle servers into or out of power saving mode \cite{lin2012dynamic}, software-based autoscaling \cite{hanafy2023carbonscaler}), and batch workload scheduling \cite{wiesner2021let,sukprasert2024limitations}. However, it's hard to adopt these approaches in commercial cloud datacenters due to cloud providers' concerns about service impact and associated revenue loss. Another approach is flexing of datacenter capabilities including cooling and DVFS \cite{chien2022beyond,gnibga2023renewable}.
One promising approach is load decoupling with energy resources as it avoids impacts on IT operation. Because of the limitations of UPS energy storage (limited capacity \cite{wang2012energy}) and diesel backup generators (carbon-heavy generation) currently deployed in datacenters, load decoupling for carbon reduction would incur additional cost. However, hyperscalers' recent actions of adding generator or energy storage to datacenters in response to requirements of grid operators \cite{msftIEGasGenforDC,vaidhynathan2025vulcan,xAIMegapack25} suggest it is a practical solution.

\section{Problem and Approach}
\label{sec:problemAndApproach}
\label{sec:problem}

\begin{figure}[h]
    \centering
    \includegraphics[width=\columnwidth]{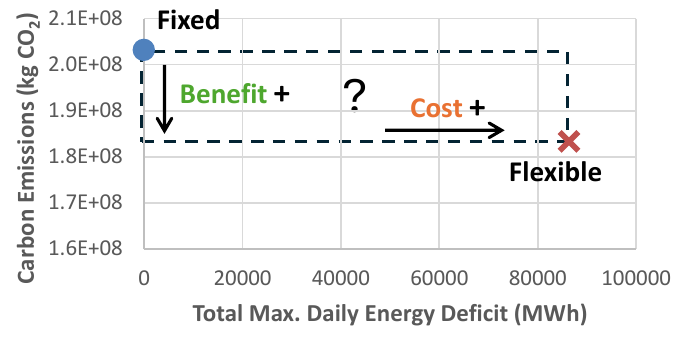}
    \caption{Datacenter flexibility reduces grid carbon emissions but realizing has a cost, forming a trade-off space to explore.}
    \label{fig:carbon_energy_tradeoff_problem}
\end{figure}

\paragraph{\textbf{Problem}}
In power grids with significant volatile renewable generation, flexing datacenters' grid load to match renewable generation can reduce grid carbon emissions, dispatch cost, etc. This can mitigate the grid challenges that DC growth creates. Figure \ref{fig:carbon_energy_tradeoff_problem} illustrates a cost-benefit space.  At the upper left are 
fixed load DCs (current practice).  At the lower right are flexible load DCs, controlled by grid dispatch to maximize grid welfare.  Modeling results show that flexibility in DCs can reduce grid carbon emissions by 10\% in this wind-dominated grid. 
Achieving this benefit requires significant flexibility in DC load. One measure of daily flexibility, the total maximum daily energy deficit (defined in Section \ref{subsec:loadVarProp}), is  86,400 MWh (21\% of daily power consumption) to achieve it.

Achieving grid load variation simultaneously with stable DC capacity, e.g. decoupling, is difficult and can be expensive. Decoupling of DC grid load and power capacity can be achieved with a collection of co-located energy resources (e.g. energy storage, generators). If DC capacity variation is allowed, it can be harmful to DC productivity \cite{zhang2021scheduling}, which incurs high direct and indirect costs \cite{dcOutageCost2016}. Therefore, We focus on the decoupling approach. Decoupling allows compute performance unaffected while the grid load is flexed, but it also incurs costs (e.g. equipment, operation) additional to current datacenter infrastructure. Naturally, a key question for load decoupling is: \textbf{are the benefits worth the costs? and does the answer differ from DC and grid perspectives?} Subsidiary research questions include:
\begin{itemize}
    \item What is the cost-benefit trade-off?
    \item How to distribute a total decoupling budget across datacenters for maximum potential benefits?
    \item How to efficiently manage the distributed decoupling capability?
    \item How to attribute the benefits to incentivize support for load adaptation?
    \item Does the grid generation composition  affect these answers? 
\end{itemize}

\paragraph{\textbf{Approach}}
%There is evidence that hyperscalers are considering adding energy resources such as battery storage and generators to meet the load flexibility requirement of grid operators \cite{vaidhynathan2025vulcan,msftIEGasGenforDC}. This enables them to flex the grid load of datacenter with internal power capacity kept constant, avoiding the higher cost from service disruption.

To understand the need for decoupling for load flexibility, we first define the key dimensions of load profiles that capture the power and energy capacity required.  These dimensions characterize the decoupling needs to maximize grid benefits, and we use them as starting points in the trade-off space (e.g. the cross in Figure \ref{fig:carbon_energy_tradeoff_problem}).

With the formalized needs, designing DC load decoupling has two phases: distribution and management.  First, distribution decides the static decoupling capacity at each datacenter given a total budget.  Second, management controls the distributed capacity to flex datacenter grid load. These two phases together produce the costs (capital, operational) and benefits (operational) of load decoupling.  Varying the total decoupling budget frames the trade-off.

We explore the design space for distribution and management of datacenter load decoupling with the goal of achieving high datacenter and grid benefits at low cost. In the following sections, we proceed as follows:
\begin{enumerate}
    \item Analyze the power and energy needs of decoupling given targeted datacenter grid load and capacity profiles.
    \item Propose several distribution and management approaches for decoupling  with desirable properties (e.g. effectiveness, datacenter autonomy).
    \item Evaluate their costs and benefits from the perspectives of both datacenters and the grid.
\end{enumerate}

\section{Designing Datacenter Load Decoupling}% for the Grid}
%Datacenter Grid Load Adaptation with Decoupling}
\label{sec:loadAdaptation}

\subsection{What is the Local Decoupling Need?}
%\subsection{Decoupling-related Load Variation Properties}
\label{subsec:loadVarProp}
Datacenter operators typically aim to achieve constant, high resource capacity (machines and power) utilization. 
This maximizes capital efficiency, and productive output for very expensive datacenters \cite{techCapexGrowth2025May,stargate500B}.  The corresponding workload management, along with hardware constraints such as poor server energy proportionality, produces stable (nearly constant) datacenter grid loads. 
If to help the grid, datacenter power capacity were varied with grid dynamics, that would in many cases harm compute efficiency and associated service revenue \cite{zhang2021scheduling,xing2023carbon}.  We consider an alternative, decoupling the datacenter power capacity from its grid load with energy resources (e.g. generators and energy storage). This approach enables the best of both worlds, flexible grid load and constant datacenter power capacity, but at a cost.

We characterize local decoupling need for a datacenter by studying the power (instantaneous)
differences between targeted grid load ($gridLoad$) and datacenter power capacity ($DCPower$).  We also consider the energy (accumulated power) over time.  Formally, for any time point $t$ in interval $[t_1, t_2]$\footnote{$(x)^+=x$ if $x\ge0$, else $(x)^+=0$.}:
$$\textbf{Power surplus: }decpPow^+(t)=(gridLoad_t-DCPower_t)^+$$
$$\textbf{Power deficit: }decpPow^-(t)=(DCPower_t-gridLoad_t)^+$$
As the power difference accumulates, we have:
$$\textbf{Energy surplus: }decpEn^+(t_1,t_2)=\sum_{t=t_1}^{t_2}(gridLoad_t-DCPower_t)^+$$
$$\textbf{Energy deficit: }decpEn^-(t_1,t_2)=\sum_{t=t_1}^{t_2}(DCPower_t-gridLoad_t)^+$$
$$\textbf{Net energy: }decpEn_{net}(t_1,t_2)=decpEn^+(t_1,t_2)-decpEn^-(t_1,t_2)$$

\begin{figure}[h]
    \centering
    \includegraphics[width=\columnwidth]{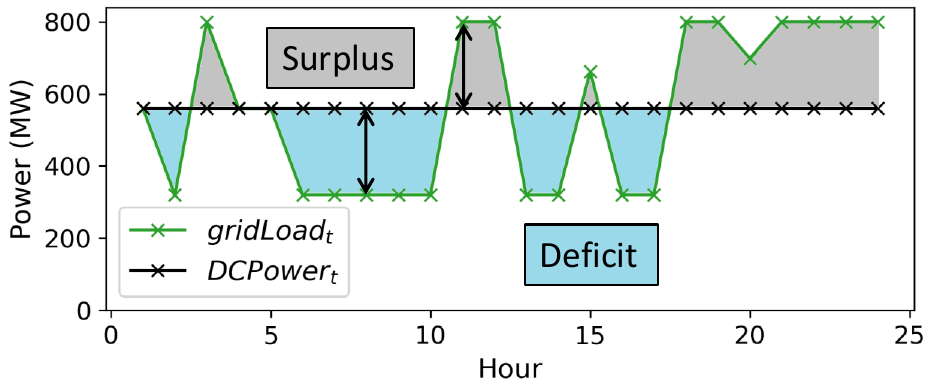}
    \caption{Flexed datacenter grid load $gridLoad_t$ produces surplus/deficit in power/energy relative to current datacenter operation practice represented by $DCPower_t$.}
    \label{fig:decpReqDemo}
\end{figure}

Aligning these terms with the power grid's daily dispatch and the targeted constant $DCPower$ profile ($DCPower_t=DCPower_{avg},\ \forall t$), over 24 hours on day $d$ (Figure \ref{fig:decpReqDemo}):
$$decpEn_d^+=\sum_{t=1}^{24}(gridLoad_t-DCPower_{avg})^+$$
$$decpEn_d^-=\sum_{t=1}^{24}(DCPower_{avg}-gridLoad_t)^+$$
Specifically, we consider flexible $gridLoad_t$ adapted to grid dynamics with $DCPower_{avg}$ as average load throughout a day, which benefits grid reliability and decarbonization \cite{ShedShiftLBNL24}. Formally,
\begin{subequations}
\label{eq:gridLoad}
\begin{gather}
    util_{min}\le gridLoad_t/DCPower_{max}\le util_{max} \label{eq:gridLoadRange}\\
    decpEn_{net}(0,t)\le 0,\ decpEn_{net}(0,24)=0 \label{eq:zeroNetEnergy}
\end{gather}
\end{subequations}
Considering the energy resources that decouples constant $DCPower$ from this flexible $gridLoad$, Eq. \ref{eq:zeroNetEnergy} enforces a circular boundary constraint across days. In the long term, the local decoupling need is defined in terms of maximum daily power deficit (aggregated power) and energy deficit (energy capacity) respectively:
$$decpPow_{max}^-=DCPower_{max}\cdot(util_{avg}-util_{min})$$
$$decpEn_{max}^-=max(\{decpEn_d^-\})$$
The decoupling need is a key determinant of the capital costs of energy resources, such as the fuel storage for a generator or the energy capacity of a battery.  
In the following sections, we explore how to distribute a total budget that constrains $\sum_{i\in DC} decpEn_{i,max}^-$ and manage local decoupling capacity ($decpEn_{i,max}^-$).

%we select a large maximum daily power deficit and more focus on varying the maximum daily energy deficit, which is closely related to ``shift potential'' in 

%\cite{ShedShiftLBNL24} and the utilization of renewable generation in the grid.

\subsection{How to Distribute Decoupling?}
We consider a global ``total decoupling'' budget, and ask how to distribute it most effectively?
We assume that the decoupling, tied to physical resources, must be done statically, and consider two distribution methods, prioritizing fairness or efficiency.

\textbf{Even distribution (EvenDist)} evenly distributes the total budget to datacenters (or proportional to their power load).  This simple method treats different datacenters equally in allocating decoupling.  This is similar to EirGrid's requirement that new datacenters need to bring generation or storage resources equivalent to their load for grid connection \cite{eirgrid23report}.

\textbf{Grid-optimized distribution (OptDist)} distributes decoupling in a fashion that 
optimizes grid metrics such as dispatch cost. 
This approach takes into account grid structure (e.g. transmission, generation location) as well as datacenter location.  For example, distributing more decoupling to datacenters  close to renewable generators or experiencing severe transmission congestion can be more beneficial to the grid.   To find OptDist distributions, we formulate a stochastic optimization problem. Formally,
\begin{subequations}
\label{eq:controlForConstCap}
    \begin{gather}
        \boldsymbol{\min} \quad 
        \sum_{d\in S}w_d\cdot dispatchCost_s \label{eq:planObj}\\
        \textbf{s.t.} \quad
        \sum_{i\in DC} decpEn_{i,max}^-\le totalDecpEn \label{eq:decpEnBudget}\\
        decpEn_{i,d}^- \le decpEn_{i,max}^-,\ \forall d \label{eq:dailyDecpEn}\\
        \text{$gridLoad$ flexibility constraints (\ref{eq:gridLoadRange}), (\ref{eq:zeroNetEnergy})},\ \forall d\nonumber \\
        \text{grid DC-OPF constraints},\ \forall d\nonumber
    \end{gather}
\end{subequations}
The objective is to minimize the weighted average grid dispatch cost, a metric that reflects overall grid welfare (defined in Section \ref{subsec:method}), over a set of scenarios $S$ covering season and weekday/weekend variation. Constraint \ref{eq:decpEnBudget} distributes the total budget $totalDecpEn$ to different datacenters as a static property across the scenarios. For each day $d\in S$, datacenter load adaptation produces energy deficit not more than the distributed limit (\ref{eq:dailyDecpEn}). This formulation is a special case of the optimal grid resource planning, where the resources are only added to datacenter sites to directly mitigate the negative impacts of datacenter growth. More general placement arrangements are possible \cite{xiong2015optimal,jabr2014robust}.

By varying the total budget $totalDecpEn$, we can explore the trade-off between decoupling capacity and benefits.

\begin{figure*}[ht]
    \centering
    \includegraphics[width=1.6\columnwidth]{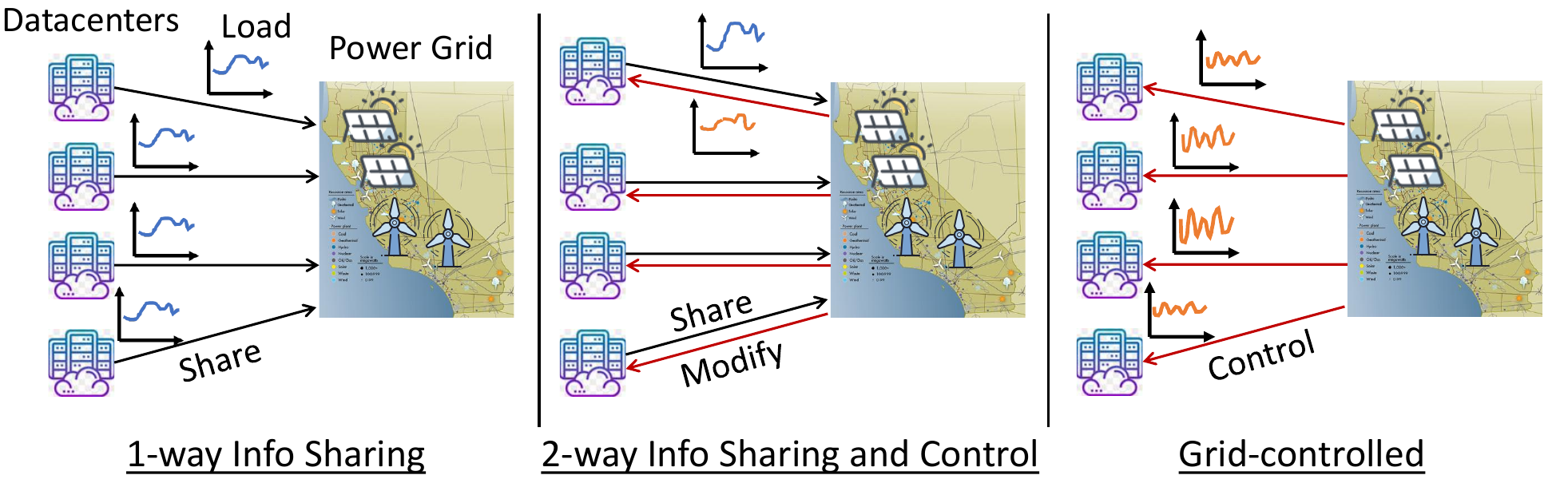}
    \caption{Decoupling Management Approaches varying in Datacenter Autonomy. PlanShare, PS-GridScale, and GridCtrl belong to the three categories respectively.}
    \label{fig:coordinationComp}
\end{figure*}

Note that another obvious way to distribute decoupling capacity would be in proportion to maximum local need.  We revisit this question in the discussion of Figure \ref{fig:siteDecpEnDist}.

\subsection{How to Manage Distributed Decoupling?}

If decoupling is owned or paid for by datacenters, then the operators may prefer to
preserve some management autonomy. This is different from GridCtrl scenario considered where the power grid controls the decoupling resources.
We frame two additional management approaches that preserve autonomy and also GridCtrl. Where datacenters have management autonomy, cooperation with the grid is also considered as it's important for effective grid carbon reduction \cite{lin2023adapting}. 

% With energy resources that correspond to $decpEn_{i,max}^-$ equipped and ${gridLoad_t}$ subject to Constraint \ref{eq:dailyDecpEn}, datacenters can flex their grid load with constant power capacity guaranteed. Because datacenters are large loads (up to multi-GWs! \cite{MetaBuildingGWDC}) that collectively account for 10\%+ or even 20\%+ of grid load in many places \cite{dominionVA23report,eirgrid23report,ercotLoadForecast}, 

%It's important to coordinate their load adaptation behaviors and grid dispatch \cite{lin2023adapting}. We consider three categories of coordination approaches varying in how $gridLoad_t$ is adapted to grid dynamics.

\paragraph{\textbf{DC-adapt with 1-way Info Sharing}} This approach extends  classical local, selfish load adaptation, sharing information about future datacenter load with the grid.  This approach preserves the autonomy of datacenters in $gridLoad_t$ while enabling grid dispatch to exploit the load information for optimization. There are a diverse set of grid metrics (e.g. power price, carbon intensity) \cite{lindberg2021guide} which datacenters can adapt grid load to based on the objective (e.g. minimizing power cost, carbon emissions), availability, and effectiveness. Specifically, we evaluate \textbf{PlanShare} \cite{lin2023adapting} that makes a 24-hour load plan adapted to locational marginal price (LMP)\footnote{Other grid metrics such as locational marginal carbon intensity are also usable as information availability evolves.}---a widely usable metric correlated with local carbon intensity---and shares that load plan with the grid day-ahead. Formally, for a datacenter's operation in a day, $\{gridLoad_t\}$ is decided by:
\begin{subequations}
\label{eq:DCSelfOpt}
    \begin{gather}
        \boldsymbol{\min} \quad 
        \sum_{t=1}^{24} LMP_t\cdot gridLoad_t \label{eq:DCSelfOptObj}\\
        \textbf{s.t.} \quad
        |gridLoad_t-gridLoad_{t-1}|\le stepSize, 2\le t\le 24 \label{eq:PlanShareStepSize}\\
        decpEn_d^- \le decpEn_{max}^- \label{eq:PlanShareDecpEn}\\
        \text{Constraints (\ref{eq:gridLoadRange}), (\ref{eq:zeroNetEnergy})}\nonumber
    \end{gather}
\end{subequations}
Datacenters optimize their own power cost, and their load plan sharing enables grid-wide benefits.  Constraint \ref{eq:PlanShareStepSize} limits large load changes that could harm the grid, where $stepSize$ is tuned based on grid configurations. We add constraint \ref{eq:PlanShareDecpEn} to ensure the load shape meets the decoupling energy limit. 

\paragraph{\textbf{2-way Info Sharing and Control}} Load variation in grid load profiles fully determined by datacenters is subject to local view and can cause grid problems even with PlanShare. For example, LMP only indicates the type of generator that will be dispatched next if load increases but doesn't indicate how much load increase it can support, which is a common issue with marginal metrics. Synchronized behaviors of increasing load due to correlated LMPs across datacenter sites can oversubscribe the renewable curtailment. In contrast, the global view and control of grid dispatch enable it to avoid such problems. We design a new coordination approach---\textbf{P(lan)S(hare)-GridScale}---that enables the grid to modify the load plans proposed by datacenters. First, datacenters make 24-hour load plans as PlanShare. And then the grid imposes scale factors $\{\alpha_{i,t}\}$ on the load deviation from constant load at different datacenters to further reduce grid dispatch cost. Formally, let $gridLoad'_{i,t}$ denote datacenter $i$'s proposed grid load at time $t$, $gridLoad_{i,t}$ denote the load finalized by the grid. Decision variables $\{\alpha_{i,t}\}$ and the following constraints ($t=1,...,24$, $\forall i\in DC$) are added to the grid dispatch:
\begin{subequations}
\label{eq:varScale}
    \begin{gather}
        gridLoad_{i,t} = DCPower_{i,avg} + \alpha_{i,t}(gridLoad'_{i,t} - DCPower_{i,avg})\\
        0\le \alpha_{i,t}\le 1
    \end{gather}
\end{subequations}

PS-GridScale combines datacenter autonomy and global control of the grid. A property of PS-GridScale favorable to datacenters is that the magnitude of decoupling power after grid modification doesn't exceed the initial proposal.

\paragraph{\textbf{Grid-controlled Adaptation}} Datacenter load levels are decided by the grid dispatch, respecting the flexibility constraints (e.g. dynamic range of $gridLoad_t$).  This is an idealized approach, not practiced in today's power grids.\footnote{The only time this happens to a limited degree is during grid emergencies to avoid grid collapse \cite{ercot21outage}.}  In this paper, we focus on a variant (\textbf{GridCtrl}) targeting the mismatch between datacenters' fixed grid load and variable renewables, which regularly adjusts DCs' grid load to maximize grid benefits:
\begin{subequations}
\label{eq:gridCtrlOpt}
    \begin{gather}
        \boldsymbol{\min} \quad 
        dispatchCost \label{eq:GridCtrlObj}\\
        \textbf{s.t.} \quad
        decpEn_d^- \le decpEn_{i,max}^- \\
        \text{Constraints (\ref{eq:gridLoadRange}), (\ref{eq:zeroNetEnergy})} \nonumber \\
        \text{grid DC-OPF constraints} \nonumber
    \end{gather}
\end{subequations}
GridCtrl decides datacenters' $\{gridLoad_{i,t}\}$ as additional decision variables in the DC-OPF problem (described in Section \ref{subsec:method}). This approach provides reference of maximum grid benefits but might incur high costs on the datacenter side. Similar pilot projects such as ERCOT's ``controllable load resources'' program \cite{LFLTF} are ongoing in power grids to facilitate grid connection of large loads.

\section{Evaluation}
\label{sec:evaluation}
%% methodology goes here?
\subsection{Experimental Setup}
\label{subsec:method}

\subsubsection{Grid Simulation}
\label{subsubsec:gridSim}
Our grid simulation is based on a reduced California power grid (CAISO) model from \cite{papavasiliou2013multiarea} with 225 buses (more details in Appendix \ref{appendix:gridSimDetails}). We update the generation cost of conventional generators with the latest EIA annual data \cite{eiaFuelCost}, and adding wind generation (scale existing sites) and solar generation (add new sites \cite{caiso2023transPlan,yuan2020developing}) as appropriate to create the grid types discussed below.

Because renewable penetration\footnote{Renewable penetration is defined as the ratio of renewable generation to grid demand.} has already exceeded 60\% in Germany, Peru, and Canada (and many others), and is the goal of many grids in the 2030s \cite{eirgrid23report,California-rps-90,iea2024renewables}, we assume that level of renewables and vary mix.
We study three grid types: (1) 60\% wind, 0\% solar (\textbf{wind-dominated}); (2) 30\% wind, 30\% solar (\textbf{balanced}); (3) 0\% wind, 60\% solar (\textbf{solar-dominated}).  Examples of these include EirGrid in Ireland is wind-dominated, ERCOT in Texas is projected to be balanced, and CAISO in California is solar-dominated. 

The grid model is associated with 8 profiles of base load, import, and renewable generation (wind and solar excluded) corresponding to weekday and weekend day of four seasons. For each season, we use 25 wind and solar generation profiles to model the day-to-day generation variation.

30 datacenters, each with $DCPower_{max}=800$ MW and $avgUtil=70\%$ are added at 30 random buses in the grid. The aggregated 16.8 GW average load corresponds to 38\% of grid load, matching real-world grid projections for 2030--2035 \cite{utilNewDCConnRequest2024aug,eirgrid23report,dominionVA23report}.

With the attributes or profiles of grid entities as input, the direct-current optimal power flow (DC-OPF) problem \cite{huneault1991survey} is solved by the grid for generator dispatch in a day. DC-OPF minimizes the dispatch cost (generation cost + curtailment/load shedding penalties), subject to balancing, transmission, and generator constraints. We attach the full formulation in Appendix \ref{appendix:gridSimDetails} for reviewer convenience.

\begin{table}[h]
\caption{Summary of Power Grid and Datacenter Setup}
\label{tab:experimentSetup}
\begin{center}
\begin{tabular}{p{75pt}|p{125pt}}
  \hline
  %Attribute & Value(s)\\
  %\hline
  \multicolumn{2}{l}{Power Grid (based on CAISO topology)}\\
  \hline
  Base Load & 27, 289 MW on average\\
  (Wind\%, Solar\%) & (60, 0), (30, 30), (0, 60)\\
  \hline
  \multicolumn{2}{l}{Datacenters (30 sites)}\\
  \hline
  $DCPower_{max}$ & 800 MW\\
  $DCPower_{avg}$ & 560 MW ($util_{avg}=70\%$)\\
  $[util_{min}, util_{max}]$ & [40\%, 100\%]\\
  \hline
\end{tabular}
\end{center}
\end{table}

\subsubsection{Decoupling Settings}
The dynamic range for  $gridLoad_t$ is  $[40\%,100\%]$, enabling large potential benefits. This range defines a maximum power deficit/surplus of 240 MW. For total decoupling budget, $\sum_{i\in DC} decpEn_{i,max}^-$, we consider a range from 0 to 100\% of the maximum need, exploring the full range.

%with GridCtrl and no total budget constraint provides a reference. We scale it by 0--1 to model the transition from a tight to a sufficient budget.

We construct a dataset of 200 days (8 day types * 25 wind/solar scenarios), splitting it into 4:1 for creating the optimized distribution (OptDist) and evaluating management  respectively.

\subsubsection{Implementation}
The distribution and management of load decoupling and the grid DC-OPF are implemented as linear programs (LPs) with Julia/JuMP \cite{dunning2017jump}. We solve the LPs with Gurobi solver \cite{gurobi}. The largest LP, stochastic optimization for OptDist takes about 30 minutes to solve on a 16-core laptop. 

Specifically, the process of simulating load adaptation initiated by datacenters is:
\begin{enumerate}
    \item Initial day-ahead dispatch: solve grid DC-OPF with datacenters as fixed loads, which produces grid metrics that guide load adaptation.
    \item Each datacenter solves its decoupling management LP to propose $\{gridLoad_{i,t}\}$. For PS-GridScale, the grid collects the proposed load of all datacenters and solves a global optimization to determine $\{\alpha_{i,t}\}$ applied to different datacenters.
    \item Actual dispatch: solve grid DC-OPF with adapted $\{gridLoad_{i,t}\}$. We regard the output as realized grid metrics and report it in evaluation.
\end{enumerate}
With GridCtrl decoupling management, grid operation is simulated by solving grid DC-OPF with solutions of GridCtrl optimization (\ref{eq:gridCtrlOpt}).

\subsubsection{Carbon Accounting for Datacenters}
In the base case where datacenters are fixed loads, each DC's carbon emissions are calculated by $\sum_t ACI_t\cdot gridLoad_{t}$ where $ACI_t$ denotes the grid average carbon intensity in the $t$-th hour. After adding datacenter flexibility with decoupling, we consider two methods to allocate the grid reduction to datacenters:

\begin{itemize}
    \item Grid metrics-based (\textbf{GM}): datacenter carbon emissions are calculated using grid average carbon intensity before and after datacenter flexibility is added.
    \item Action-based (\textbf{Act}): all the grid carbon reduction is allocated to datacenters proportional to their decoupling energy. Formally, on day $d$, datacenter $i$ is allocated:
    $$\Delta gridCarbon_{d}\cdot \frac{decpEn_{i,d}^-+decpEn_{i,d}^+}{\sum_{i\in DC} (decpEn_{i,d}^-+decpEn_{i,d}^+)}$$
    % \item Shapley value-based methods:
    % \begin{itemize}
    %     \item Weight-based: use some marginal contribution metrics (e.g. 1 DC vs. no load adaptation) as weights of different datacenters' adapted load shapes and attribute the grid carbon reduction to each datacenter. 
    %     \item Operator-based: Shapley values for several operators can be computed.
    % \end{itemize}
\end{itemize}
The major difference between these two methods is that Act allocates all the grid carbon reduction to datacenters. Such attribution is considered more equitable by many researchers \cite{werner2021pricing,bona2016attribution} as datacenter flexibility accounts for the grid carbon reduction. Both of the methods would require cooperation with the grid as they need metrics from counterfactual dispatch with no DC flexibility.

\subsubsection{Evaluation Metrics}
We assess the following metrics from the perspectives of both the grid and datacenters:
\begin{itemize}
    \item \textbf{Grid Dispatch Cost (\$):} the objective of DC-OPF and measures overall grid welfare.
    \item \textbf{Grid Carbon Emissions (kg CO$_2$)} are calculated by
    $$\sum_{t=1}^{24} \sum_f gen_{f,t}*emRate_f$$
    where $gen_{f,t}$ is generation from fuel $f$ in the $t$-th hour and $emRate_f$ is fuel $f$'s emission rate.
    \item \textbf{Datacenter Carbon Emissions (kg CO$_2$):} grid carbon emissions allocated to datacenters. Calculated with grid average carbon intensity (fixed-load) and either GM or Act allocation (flexible). Assuming hyperscalers need to purchase high-quality carbon credits to offset carbon emissions, they can be monetized at \$280/metric ton CO$_2$ (direct air capture \cite{googleCarbonCapture24}).
    \item \textbf{Datacenter Power Cost (\$):} the cost of purchasing power from the wholesale power market calculated based on locational prices (LMPs).
    \item \textbf{Decoupling Cost (\$):} the cost of implementing decoupling. We consider Li-ion battery energy storage (BES), a mature and widely-usable technology, and calculate its total cost of ownership (TCO) based on its static capacity and actual usage needed for decoupling (details in Appendix \ref{appendix:BESTCO}). It can be substituted by advancing cheaper storage technologies such as gravity energy storage \cite{viswanathan2022}. 
\end{itemize}
The metrics with fixed-load datacenters (current practice) are the baseline when examining the impacts of datacenter flexibility and decoupling. For each metric, we report the average across the day scenarios with weekday/weekend day weighed by 5:2.

\subsection{Distributing Decoupling}
\label{subsec:distribEval}
\begin{figure}[h]
    \centering
    \includegraphics[width=\columnwidth]{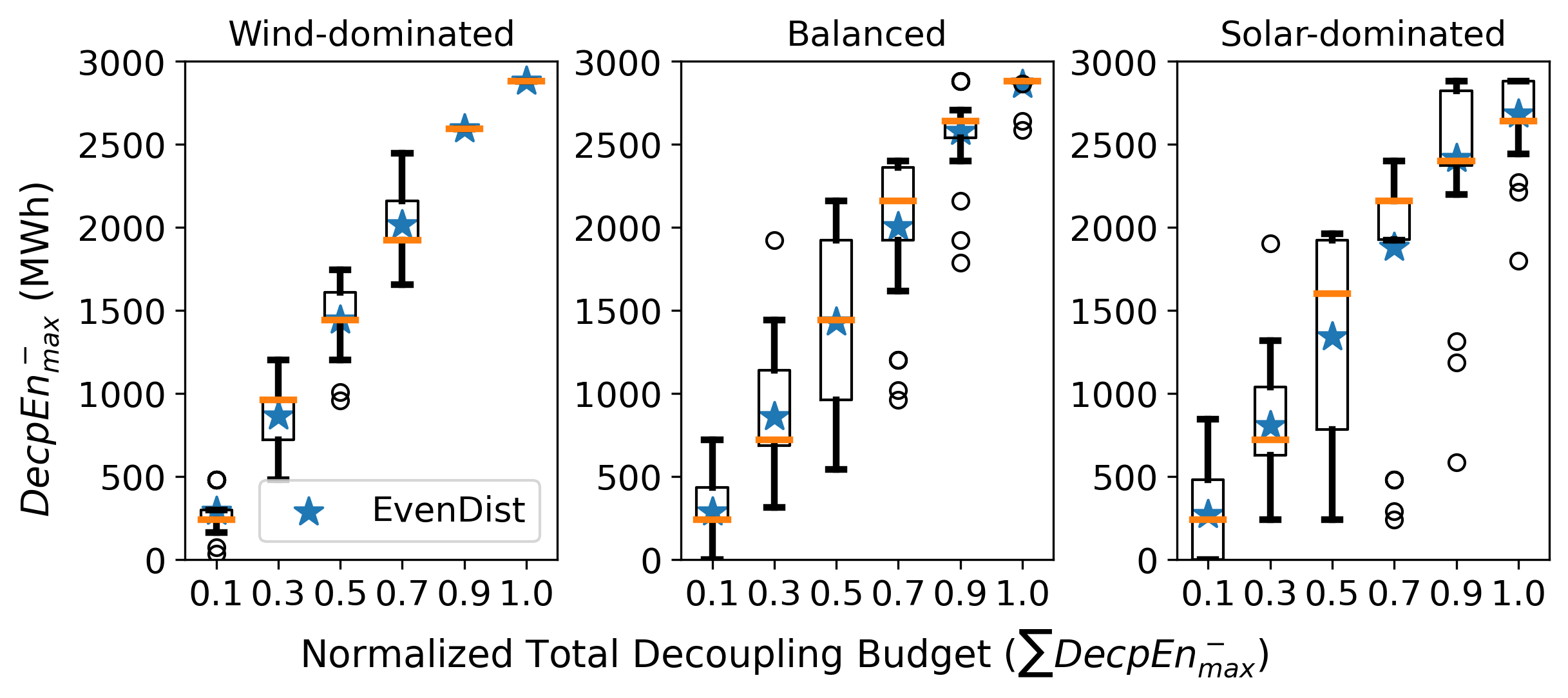}
    \caption{Decoupling Capacity Distributed to Various Datacenters by EvenDist (star) and OptDist (whiskers).}
    \label{fig:siteDecpEnDist}
\end{figure}

To study distribution of decoupling, we consider two distribution approaches: EvenDist (equal) and OptDist (grid-optimized), and vary the total decoupling budget.
We present results in terms of the normalized total budget (normalized to $\sum_{i\in DC}decpEn_{i,max}^-$ produced by GridCtrl without the budget constraint). Figure \ref{fig:siteDecpEnDist} shows the distribution results, with the stars representing the values for EvenDist (identical for all DCs), and the whiskers showing the distributions of $decpEn_{i,max}^-$ across DCs for OptDist.

The spread of the whiskers around the star reflects the difference between EvenDist and OptDist.  In most cases, an uneven distribution is needed to achieve maximum grid benefits. The distribution skew increases with more solar generation, because of its temporal correlation (daytimes), leading to transmission congestion. The top-bottom (p95-p5) ratio can be as large as 6--8x, so if datacenters had to pay for their own decoupling, some would have much larger bills. Fundamentally, the skew reflects site heterogeneity due to temporal correlation and transmission congestion, both of which grow as renewable fraction increases.

\subsection{Decoupling Impact and Benefits}

\begin{figure}[h]
    \centering
    \begin{subfigure}{.48\columnwidth}
        \centering
        \includegraphics[width=\columnwidth]{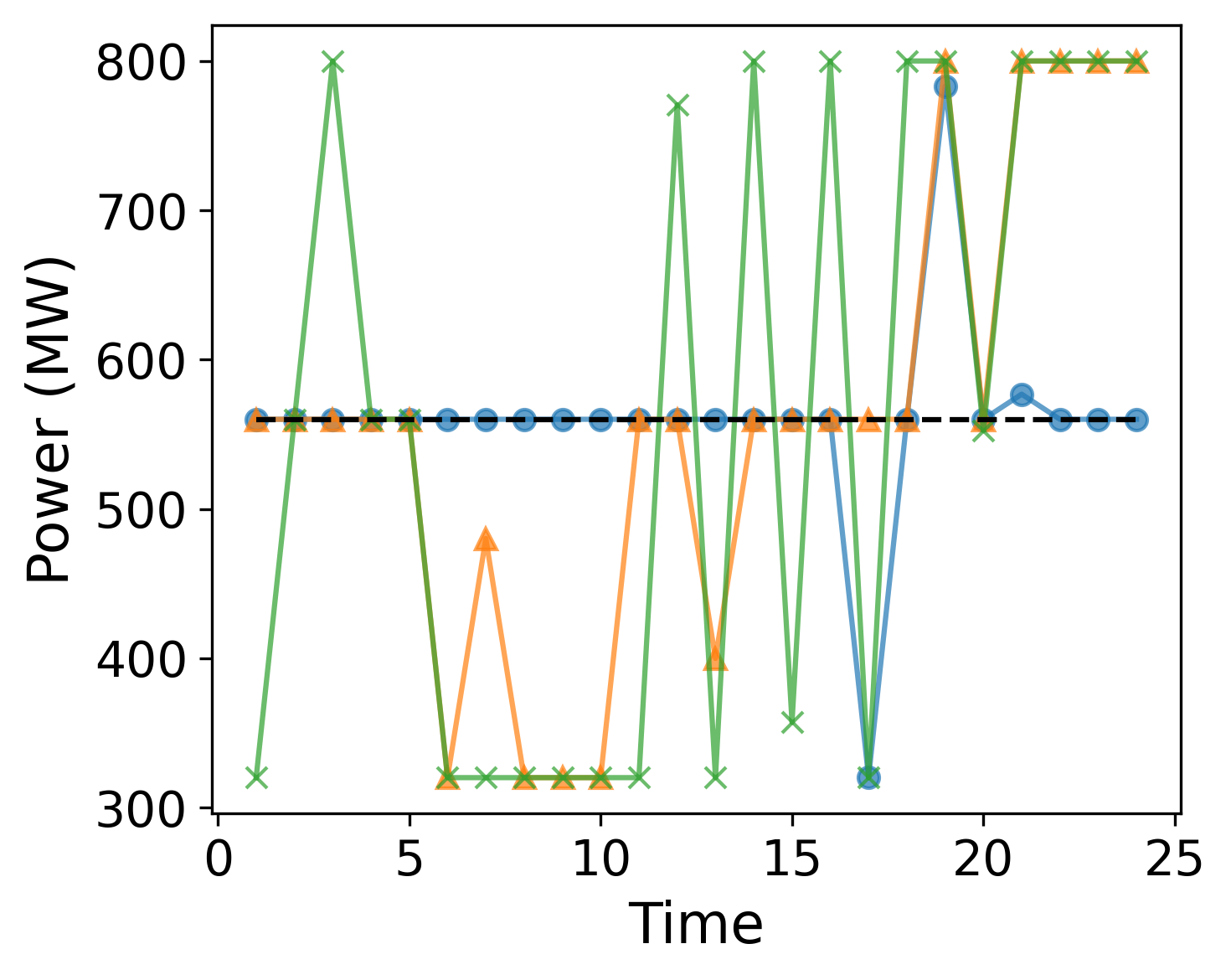}
        \caption{Wind-dominated Grid.}
    \end{subfigure}
    \begin{subfigure}{.48\columnwidth}
        \centering
        \includegraphics[width=\columnwidth]{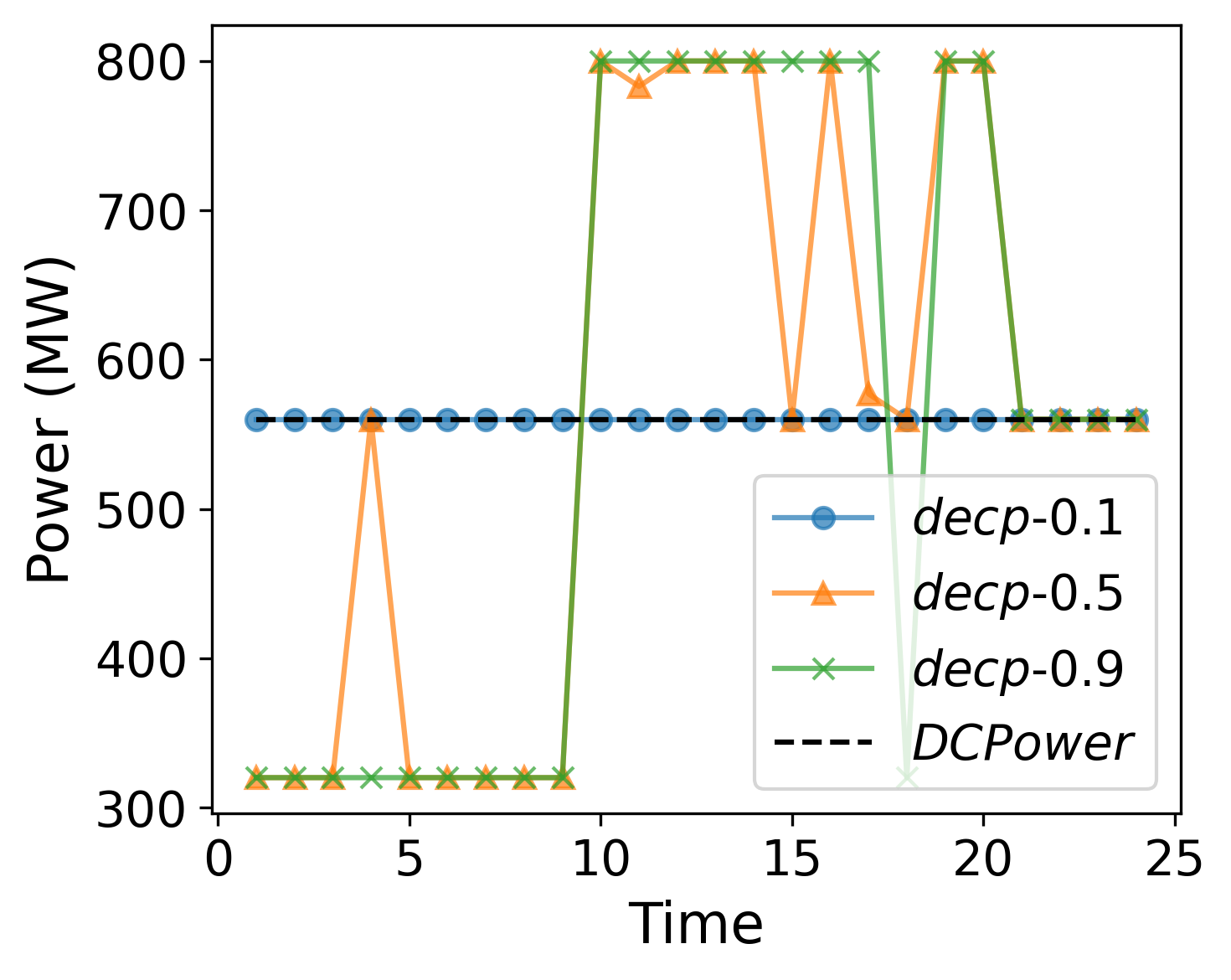}
        \caption{Solar-dominated Grid.}
    \end{subfigure}
    \caption{Datacenter Grid Load Variation with Various Decoupling Capacities (OptDist).}
    \label{fig:gridLoadVarScaleExample}
\end{figure}

We illustrate the impact of decoupling on both datacenter and grid dispatch in Figure \ref{fig:gridLoadVarScaleExample}.  First, the $DCPower$ lines show that stable power is delivered to all datacenters, all the time.  Next the ``decp-x'' lines show the impact of various quantities of decoupling capacity.  As decoupling capacity is increased, the grid load has more flexibility and evidently changes much more often. These are orchestrated by grid dispatch for benefits such as reduced dispatch cost.

\begin{figure}[h]
    \centering
    \begin{subfigure}{.48\columnwidth}
        \centering
        \includegraphics[width=\columnwidth]{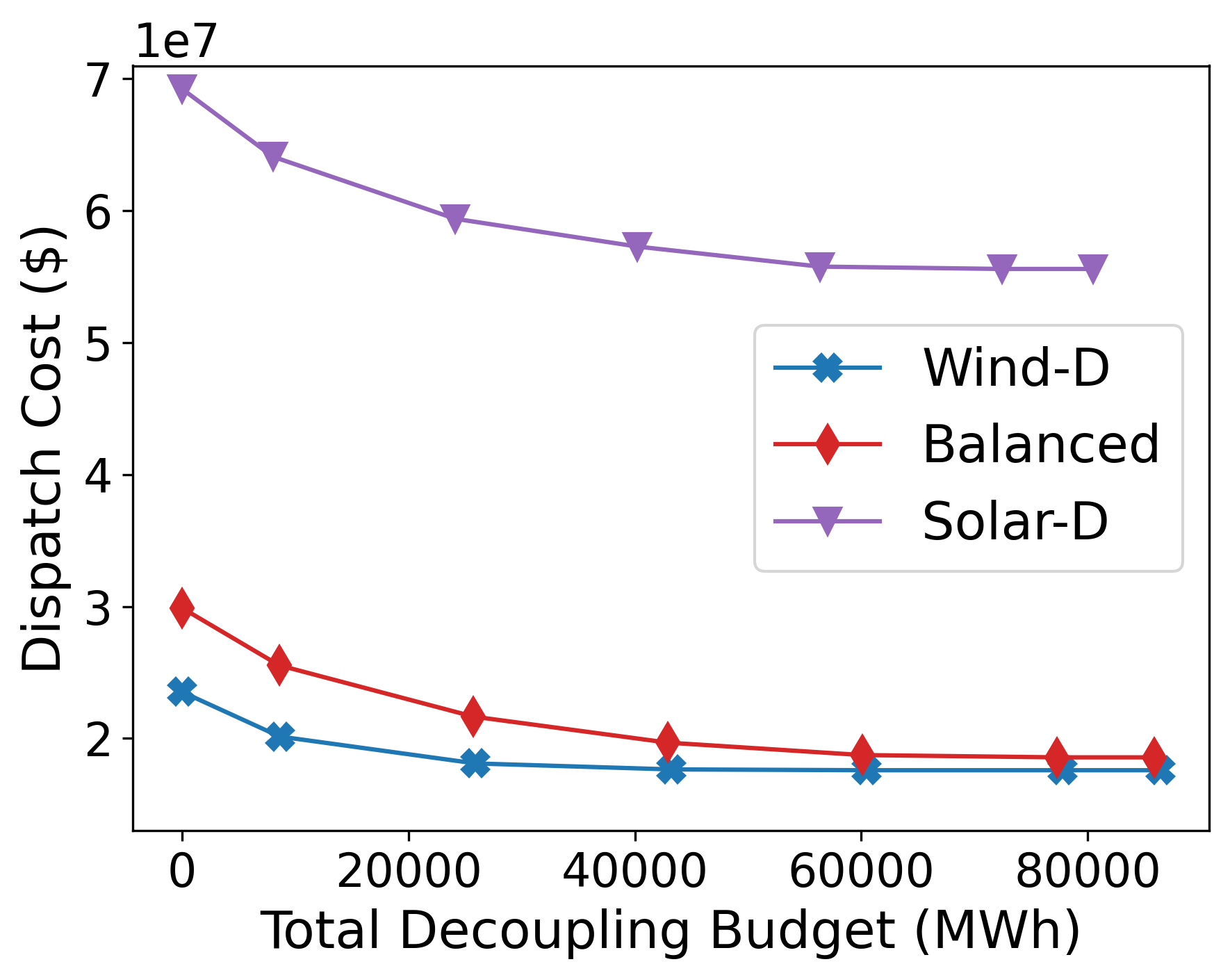}
        \caption{Dispatch Cost.}
    \end{subfigure}
    \begin{subfigure}{.48\columnwidth}
        \centering
        \includegraphics[width=\columnwidth]{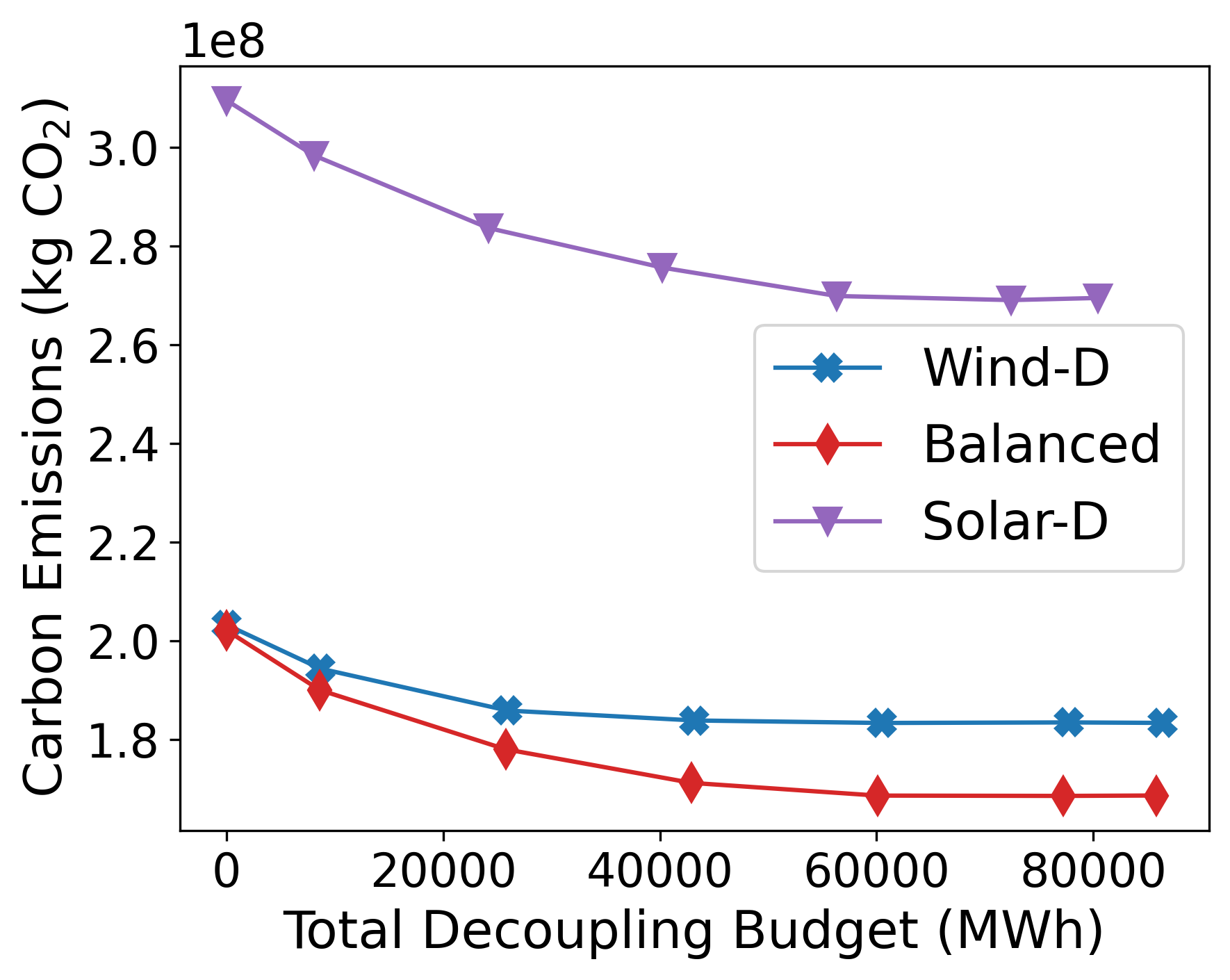}
        \caption{Carbon Emissions.}
    \end{subfigure}
    \caption{Grid Performance vs. Total Decoupling Capacity. (OptDist distribution)}
    \label{fig:gridPerf_GridCtrl}
\end{figure}

%\peterNote{Reorganized} 
Figure \ref{fig:gridPerf_GridCtrl} shows grid performance versus total decoupling capacity, starting from zero decoupling. As DC grid load flexibility increases (increased decoupling budget), both the grid dispatch cost and carbon emissions are  reduced. This is because the decoupling reduces renewable curtailment and load shedding, avoiding dispatch penalties.  Carbon emissions are reduced with better renewable absorption. With maximum decoupling budgets, across the three grid types, dispatch cost is reduced by 20--38\%, and carbon emissions are reduced by 10--17\%.  There are  diminishing returns---70\% of the maximum budget captures 98--100\% of the grid benefits across grid types, implying that significant cost savings are possible with little loss in grid benefits.  Decoupling benefits are higher in the balanced and solar-dominated grids where solar generation produces a greater need for DC flexibility.

\begin{figure}[h]
    \centering
    \begin{subfigure}{.48\columnwidth}
        \centering
        \includegraphics[width=\columnwidth]{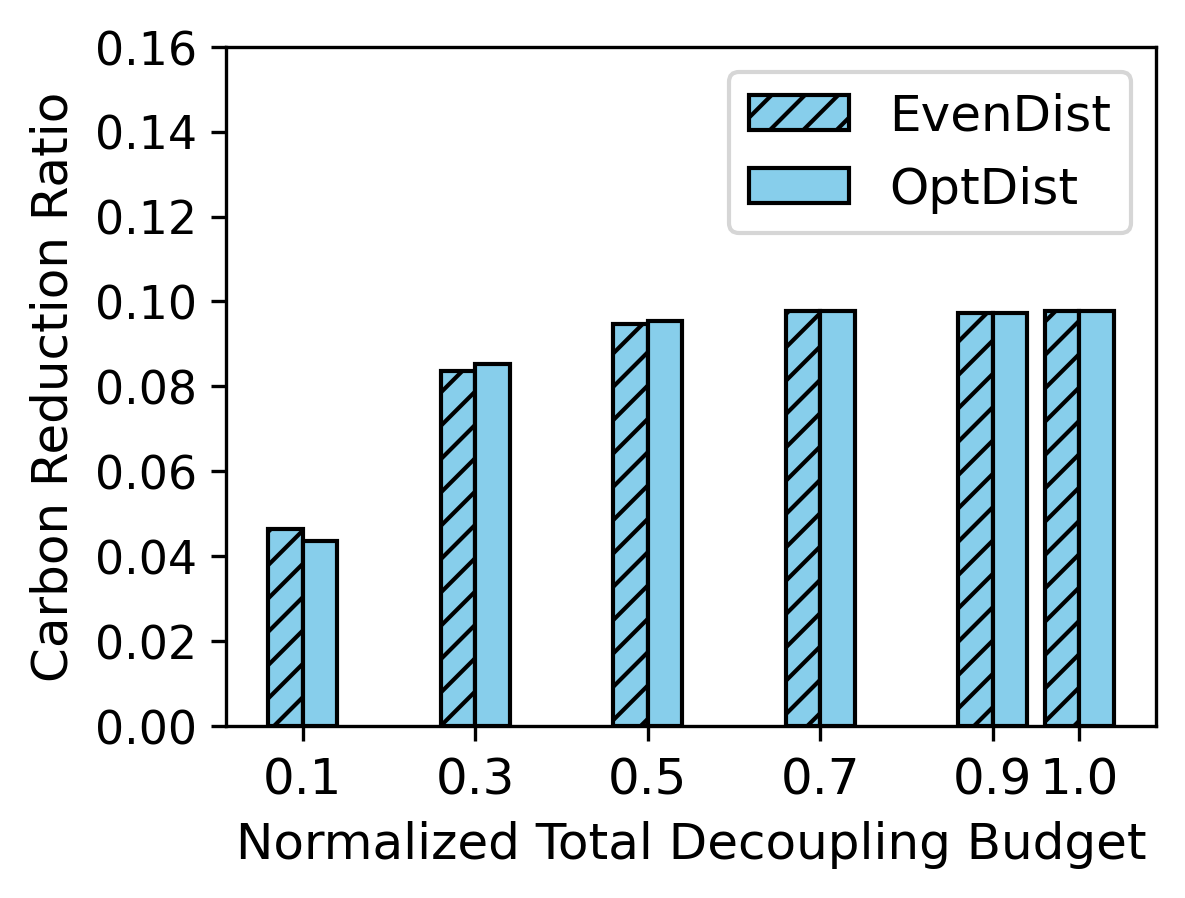}
        \caption{Wind-dominated.}
    \end{subfigure}
    \begin{subfigure}{.48\columnwidth}
        \centering
        \includegraphics[width=\columnwidth]{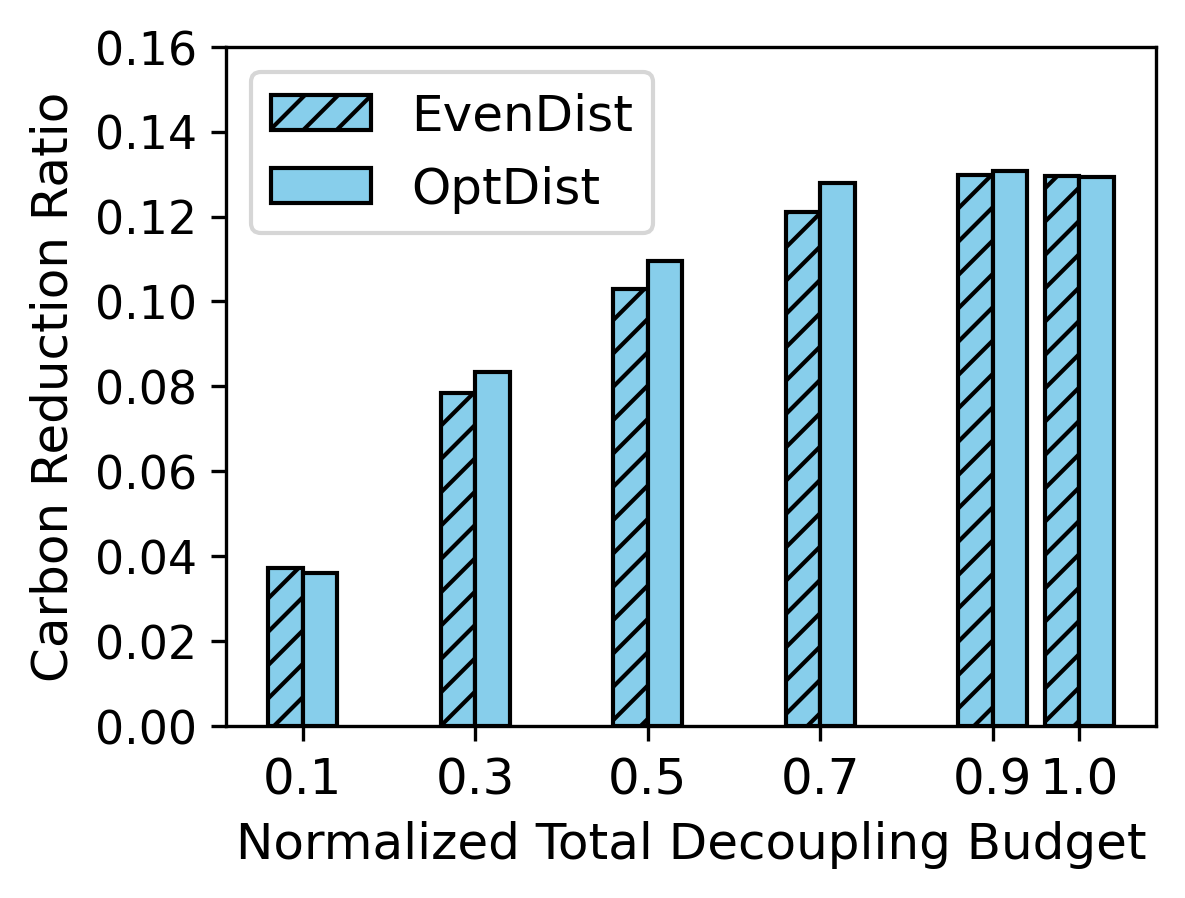}
        \caption{Solar-dominated.}
    \end{subfigure}
    \caption{Grid Carbon Reduction with Even and Optimized Distribution of Decoupling.}
    \label{fig:gridCarbonRed_distMethod}
\end{figure}

The difference in grid performance with OptDist and EvenDist is consistent with the distribution difference. In the wind-dominated grid where the skew of distribution is smaller, OptDist improves grid performance less (Figure \ref{fig:gridCarbonRed_distMethod}a); in the other two grids and with medium total budgets (0.3--0.7), the advantage of OptDist gets more evident. Figure \ref{fig:gridCarbonRed_distMethod}b shows additional grid carbon reduction of 0.7\% (2,000 metric ton CO$_2$/day). %We anticipate that the advantage will get larger when transmission capacity is more limited. 
For the following studies, the results are based on OptDist if not specified.

\subsection{Managing Decoupling Cooperatively} %for Datacenter and Grid Benefit}

We know that GridCtrl, centralized grid control of decoupling, will produce the maximum grid benefits.  However,
%because of the global optimization formulation but requires decoupling fully managed by the grid. 
if decoupling is owned or paid for by datacenters, then the operators may prefer to preserve some management autonomy.  
%This is different from the previous scenario considered where the power grid controlled all of the decoupling resources (GridCtrl).  
So we consider two schemes that retain datacenter autonomy, PlanShare and PS-GridScale, which allow datacenters to control their decoupling resources for local benefits as well.  The key question is: how do these schemes perform---for datacenter and grid? %their performance compared with GridCtrl?

\begin{figure}[h]
    \centering
    \begin{subfigure}{.48\columnwidth}
        \centering
        \includegraphics[width=\columnwidth]{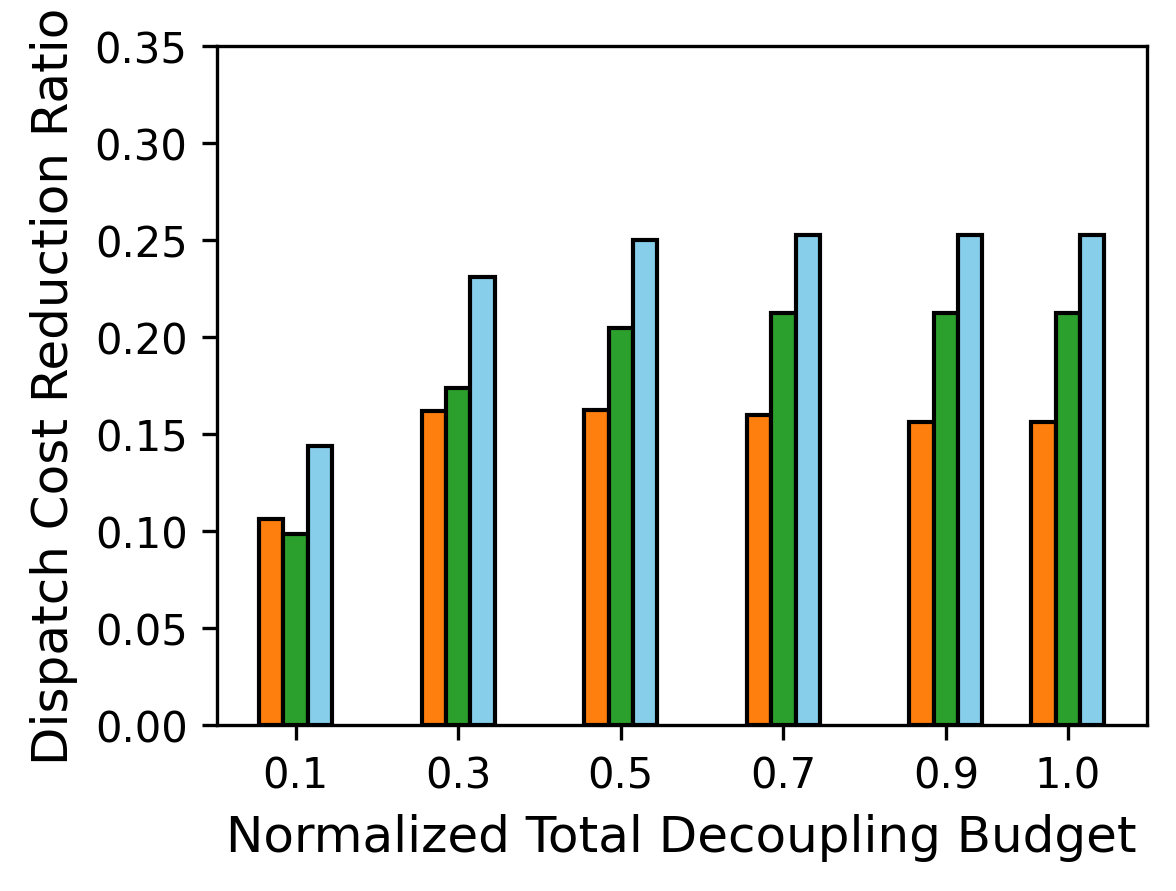}
        \caption{Wind-dominated.}
    \end{subfigure}
    \begin{subfigure}{.48\columnwidth}
        \centering
        \includegraphics[width=\columnwidth]{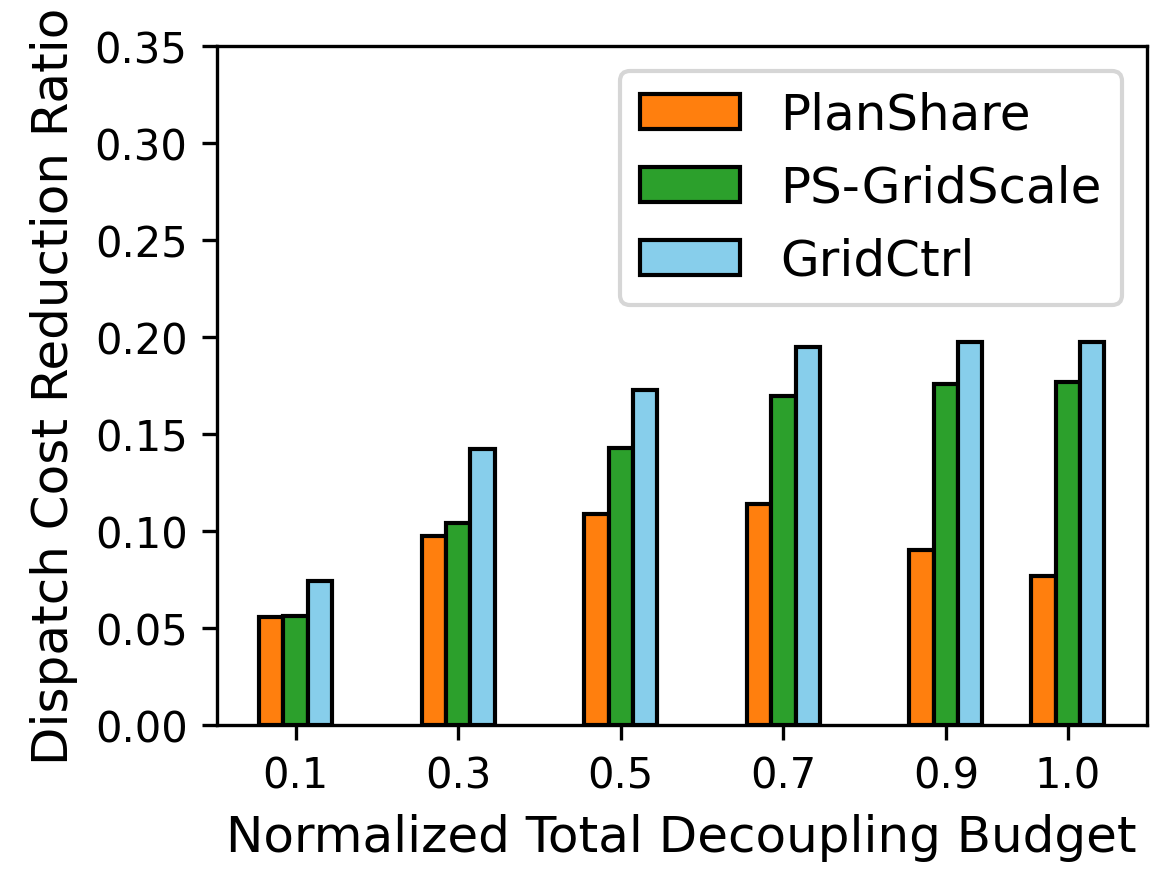}
        \caption{Solar-dominated.}
    \end{subfigure}
    \caption{Grid Dispatch Cost Reduction with Various Decoupling Management Approaches.}
    \label{fig:dispatchCostRed_management}
\end{figure}

\begin{figure}[h]
    \centering
    \begin{subfigure}{.48\columnwidth}
        \centering
        \includegraphics[width=\columnwidth]{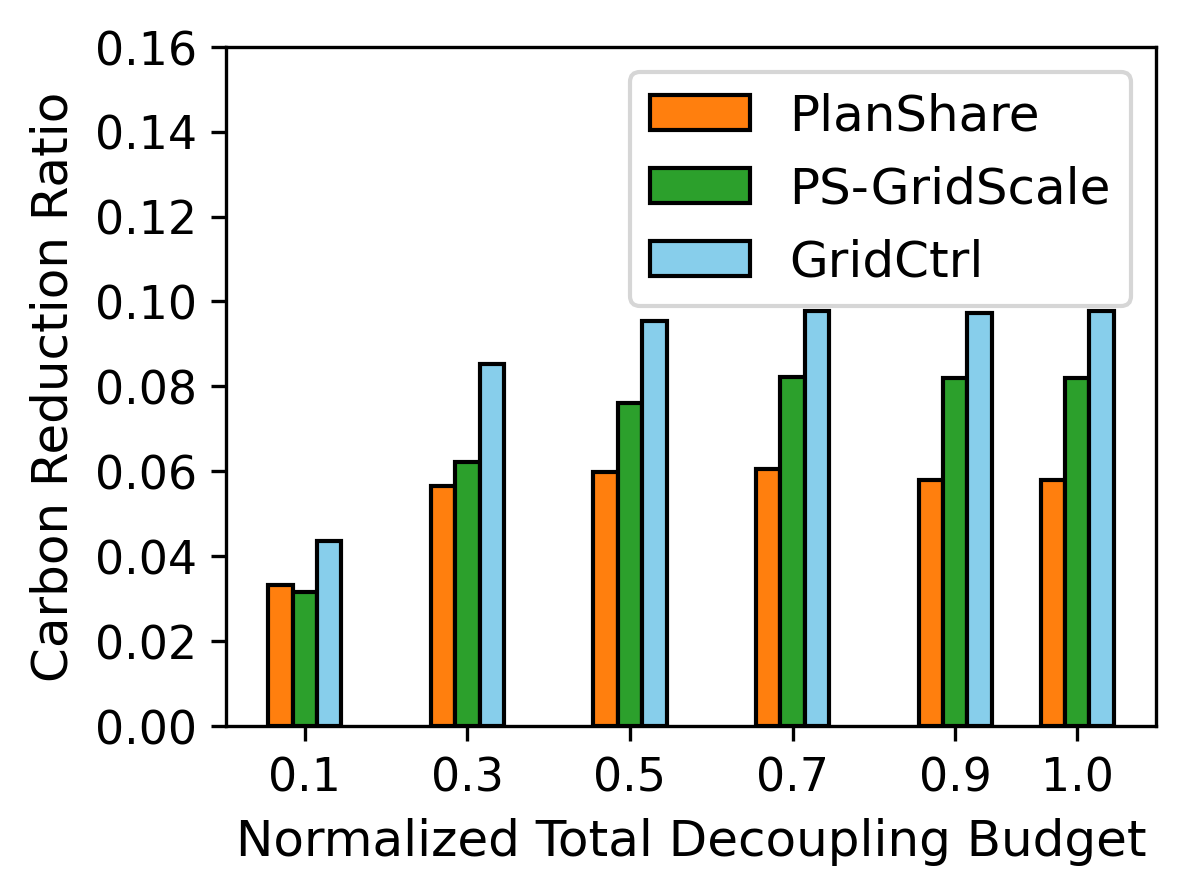}
        \caption{Wind-dominated.}
    \end{subfigure}
    \begin{subfigure}{.48\columnwidth}
        \centering
        \includegraphics[width=\columnwidth]{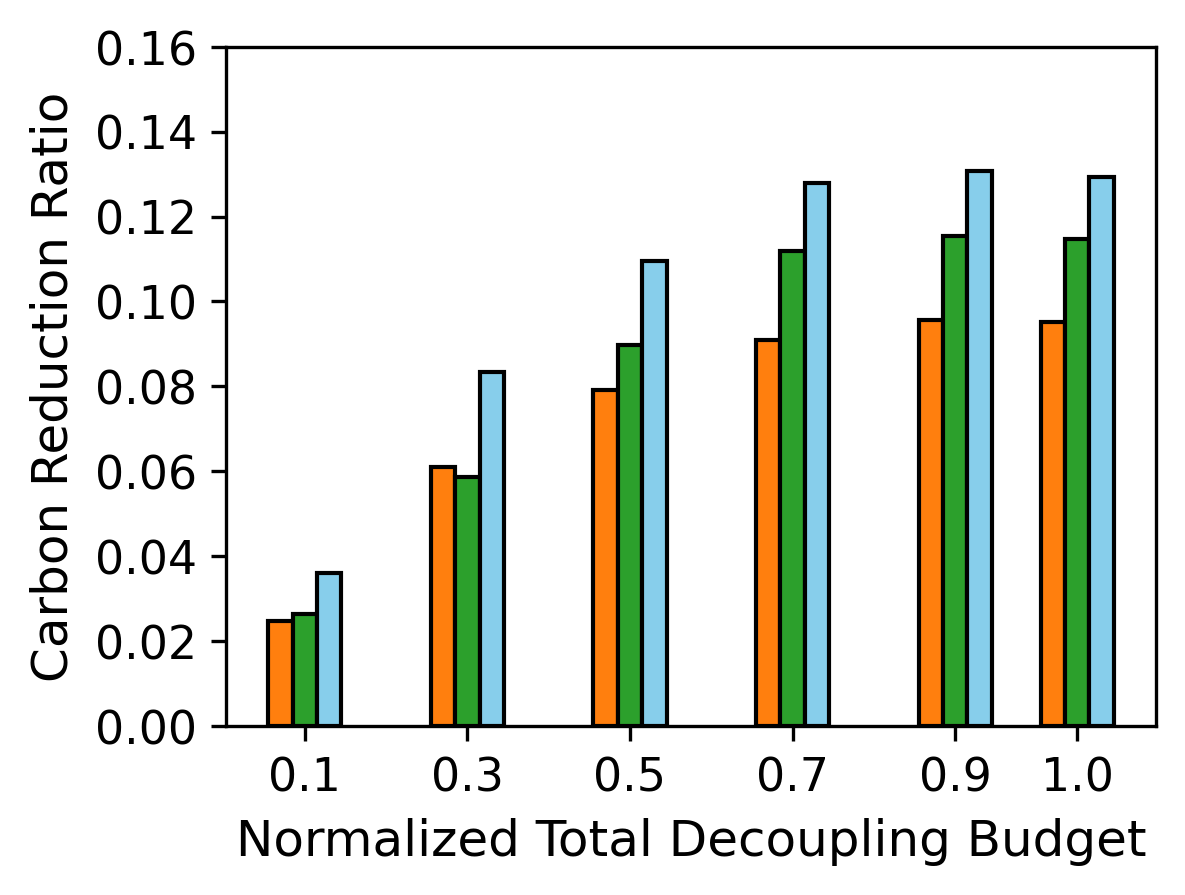}
        \caption{Solar-dominated.}
    \end{subfigure}
    \caption{Grid Carbon Reduction with Various Decoupling Management Approaches.}
    \label{fig:gridCarbonRed_management}
\end{figure}

In Figure \ref{fig:dispatchCostRed_management}, we consider grid dispatch cost.  And in Figure  \ref{fig:gridCarbonRed_management} we consider grid carbon reduction.  For both metrics (w/ OptDist distribution),
PlanShare (orange) fails to manage decoupling effectively---the benefits are generally lower that the two other schemes, and can even decrease as the total budget increases. Although still falling short of GridCtrl performance, PS-GridScale (green) consistently outperforms PlanShare, achieving 84/90\% of the dispatch cost reduction and 84/89\% of the carbon reduction with GridCtrl in the wind/solar-dominated grid. These are up to 1.6x better dispatch cost reduction and 1.4x better carbon reduction vs. PlanShare. These two approaches also see diminishing returns in decoupling capacity---most of the benefits are captured with 70\% of the maximum total budget.

\begin{figure}[h]
    \centering
    \begin{subfigure}{.48\columnwidth}
        \centering
        \includegraphics[width=\columnwidth]{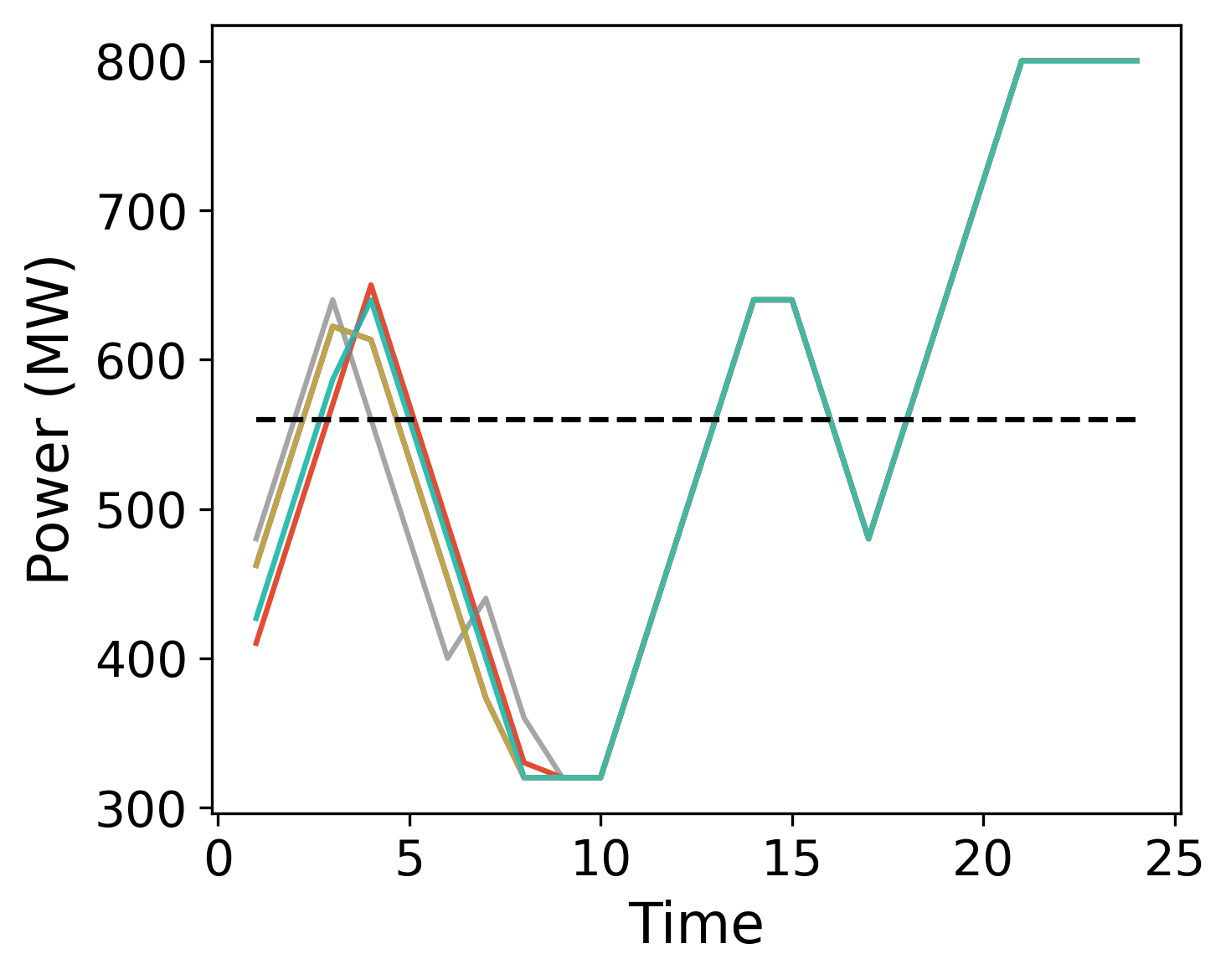}
        \caption{PlanShare.}
    \end{subfigure}
    \begin{subfigure}{.48\columnwidth}
        \centering
        \includegraphics[width=\columnwidth]{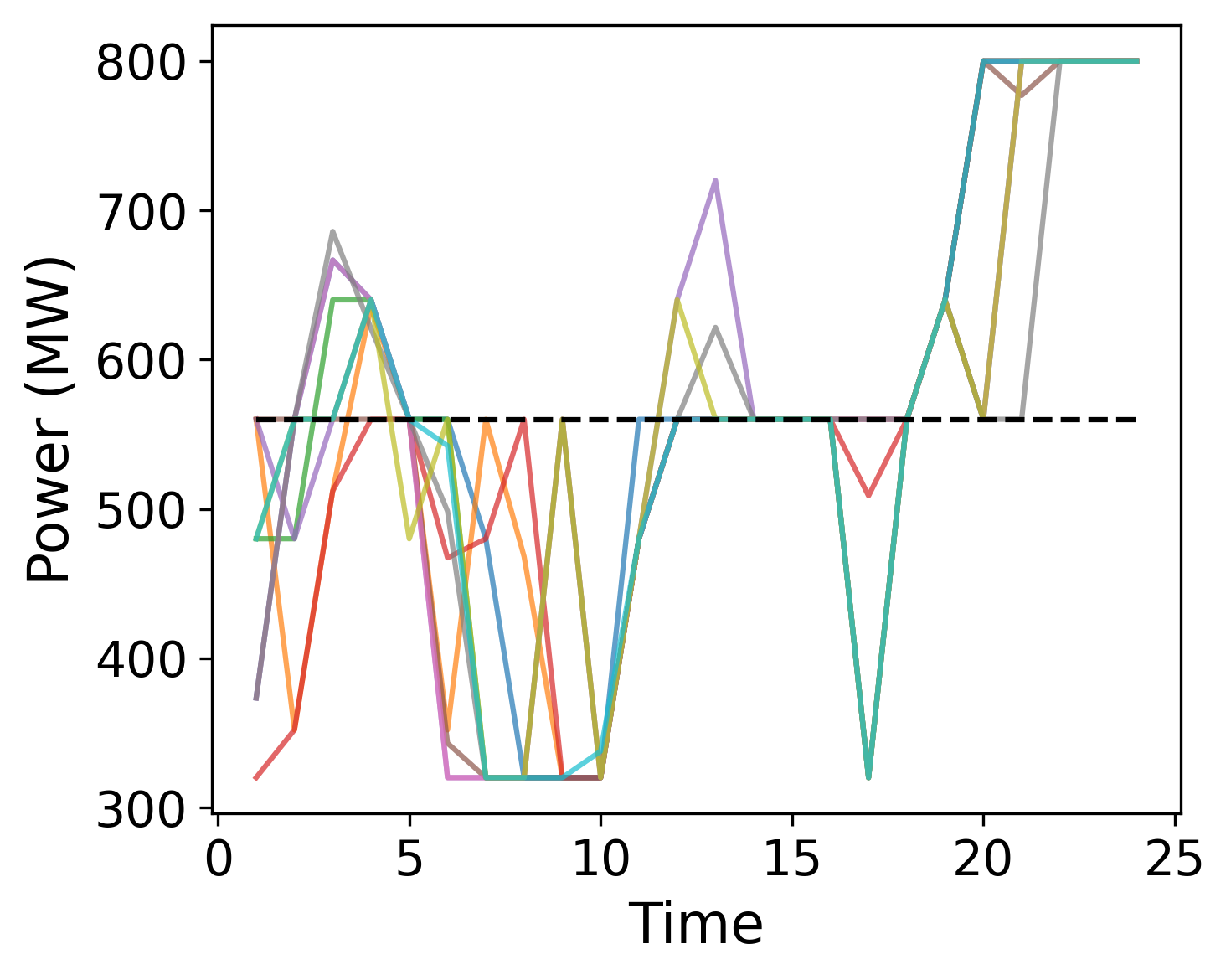}
        \caption{PS-GridScale.}
    \end{subfigure}
    % \begin{subfigure}{.33\columnwidth}
    %     \centering
    %     \includegraphics[width=\columnwidth]{figures/evaluation/DCLoad-PSandGridScale-SpringWD-0.6-315-0-32.png}
    %     \caption{Total Comparison.}
    % \end{subfigure}
    \caption{Datacenter Grid Load Examples with PlanShare and PS-GridScale Management.}
    \label{fig:DCLoad_management}
\end{figure}

To understand the performance difference between PlanShare and PS-GridScale, we examine their decoupling behaviors. Figure \ref{fig:DCLoad_management} shows the time series for grid load of 10 sampled datacenters. With PlanShare, correlated local prices (LMP) cause highly correlated grid load profiles (Figure \ref{fig:DCLoad_management}a), collectively resulting in large load variation (overshifting) that disrupts grid operation. The disruption is reflected as increased gas generation in the wind-dominated grid and load shedding in the solar-dominated grid. With PS-GridScale, although the load profiles initially proposed are also correlated on LMPs, the introduction of grid-decided scale factors breaks the correlation and thus mitigates overshifting (Figure \ref{fig:DCLoad_management}b). The overall performance advantage demonstrates the necessity of cooperation between large flexible loads and the grid.

\begin{figure}[h]
    \centering
    \begin{subfigure}{.48\columnwidth}
        \centering
        \includegraphics[width=\columnwidth]{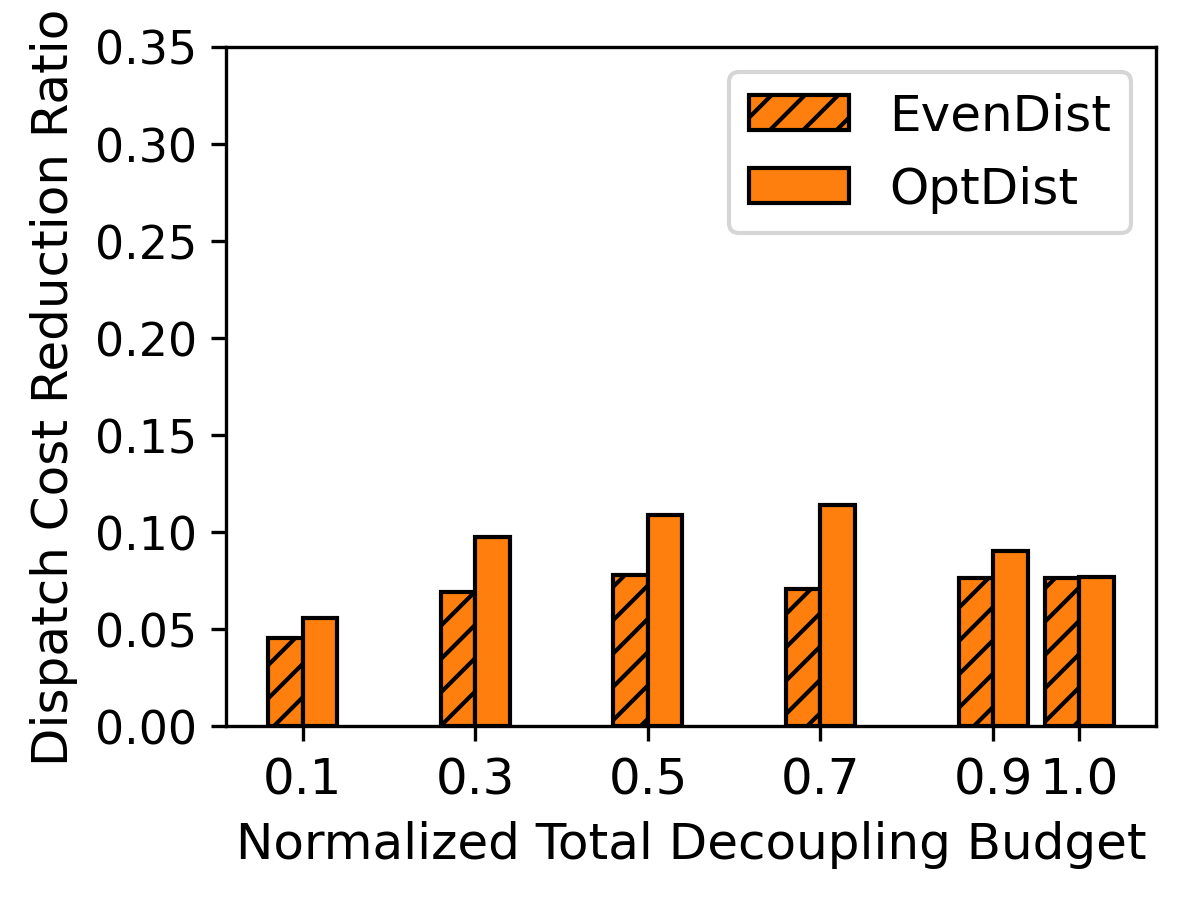}
        \caption{PlanShare.}
    \end{subfigure}
    \begin{subfigure}{.48\columnwidth}
        \centering
        \includegraphics[width=\columnwidth]{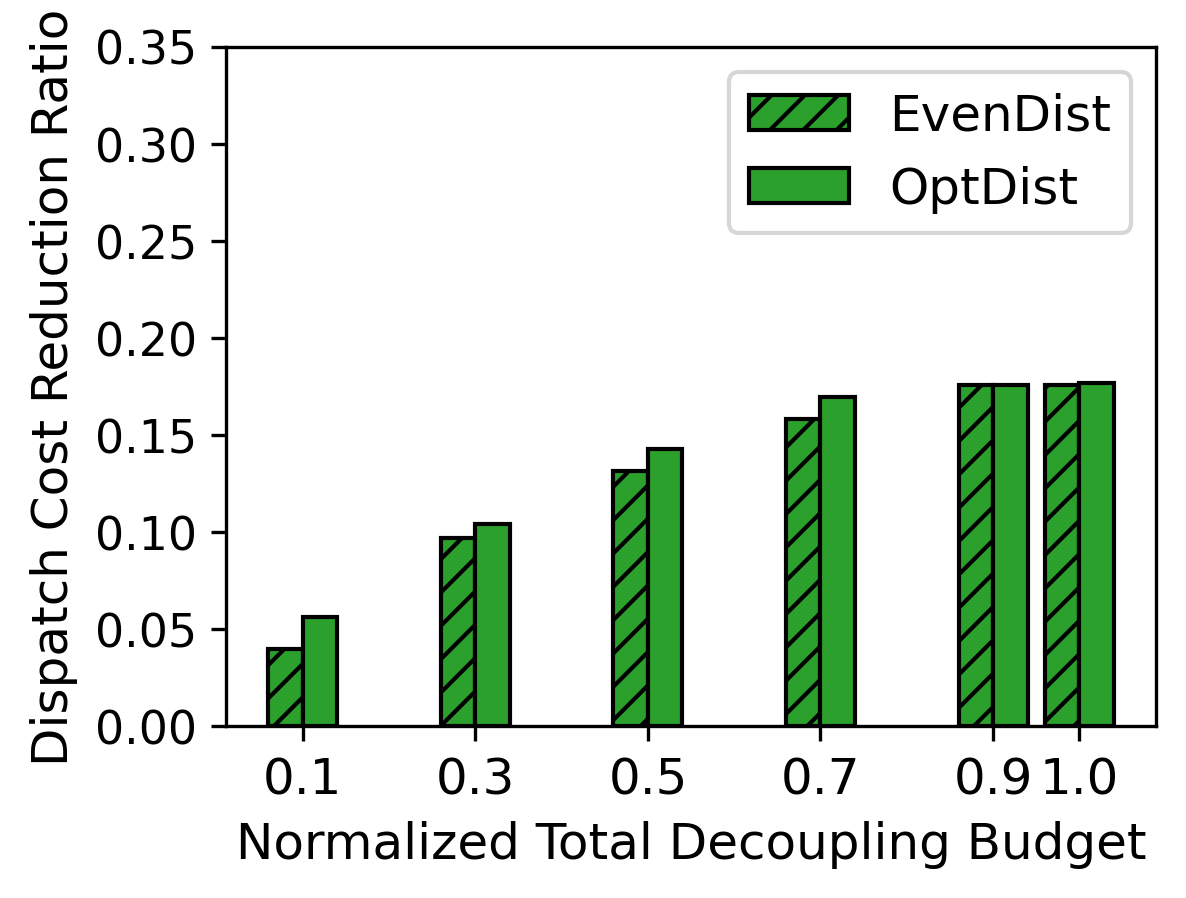}
        \caption{PS-GridScale.}
    \end{subfigure}
    \caption{Comparing Dispatch Cost for Even (EvenDist) and Optimized Decoupling Distribution (OptDist) with PlanShare and PS-GridScale. Solar-dominated Grid.}
    \label{fig:dispatchCostRed_PlanShare_distMethod}
\end{figure}

We also revisit the impact of decoupling distribution with PlanShare and PS-GridScale.  In the wind-dominated grid, the performance impact is minor, so in Figure \ref{fig:dispatchCostRed_PlanShare_distMethod} we compare grid dispatch cost in the solar-dominated grid.   PlanShare clearly benefits from OptDist, the better distribution, while PS-GridScale is robust against distribution variation.

%evidently. In this case, OptDist distributes less decoupling to locations where shifting helps the grid less, thereby limiting overshifting.  

\subsection{Economic Benefits of Decoupling}
\label{subsec:benefitVsCost}
To figure out whether decoupling is economically viable, we assess the economic benefits and and compare them with decoupling cost for datacenters and the whole grid. 

\subsubsection{Datacenter Economic Benefits from Decoupling} 
\begin{figure}[h]
    \centering
    \begin{subfigure}{.48\columnwidth}
        \centering
        \includegraphics[width=\columnwidth]{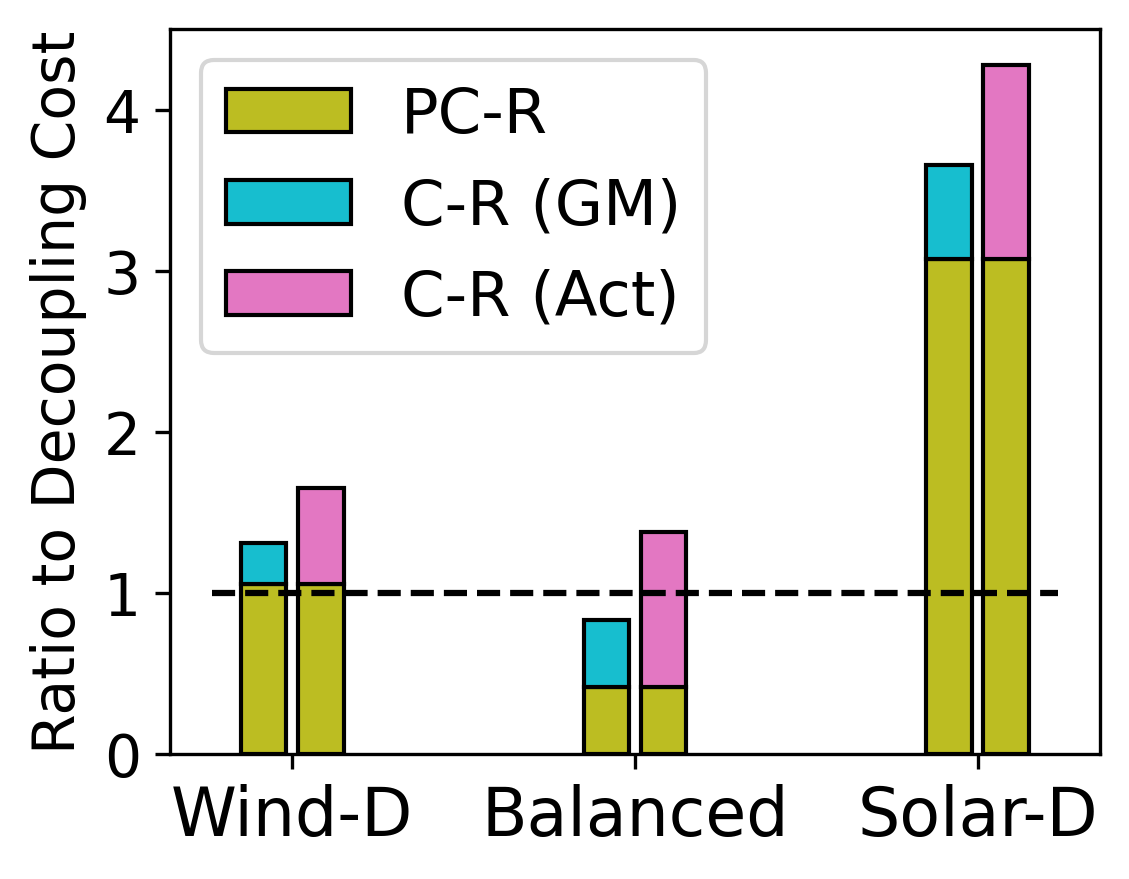}
        \caption{PS-GridScale.}
    \end{subfigure}
    \begin{subfigure}{.48\columnwidth}
        \centering
        \includegraphics[width=\columnwidth]{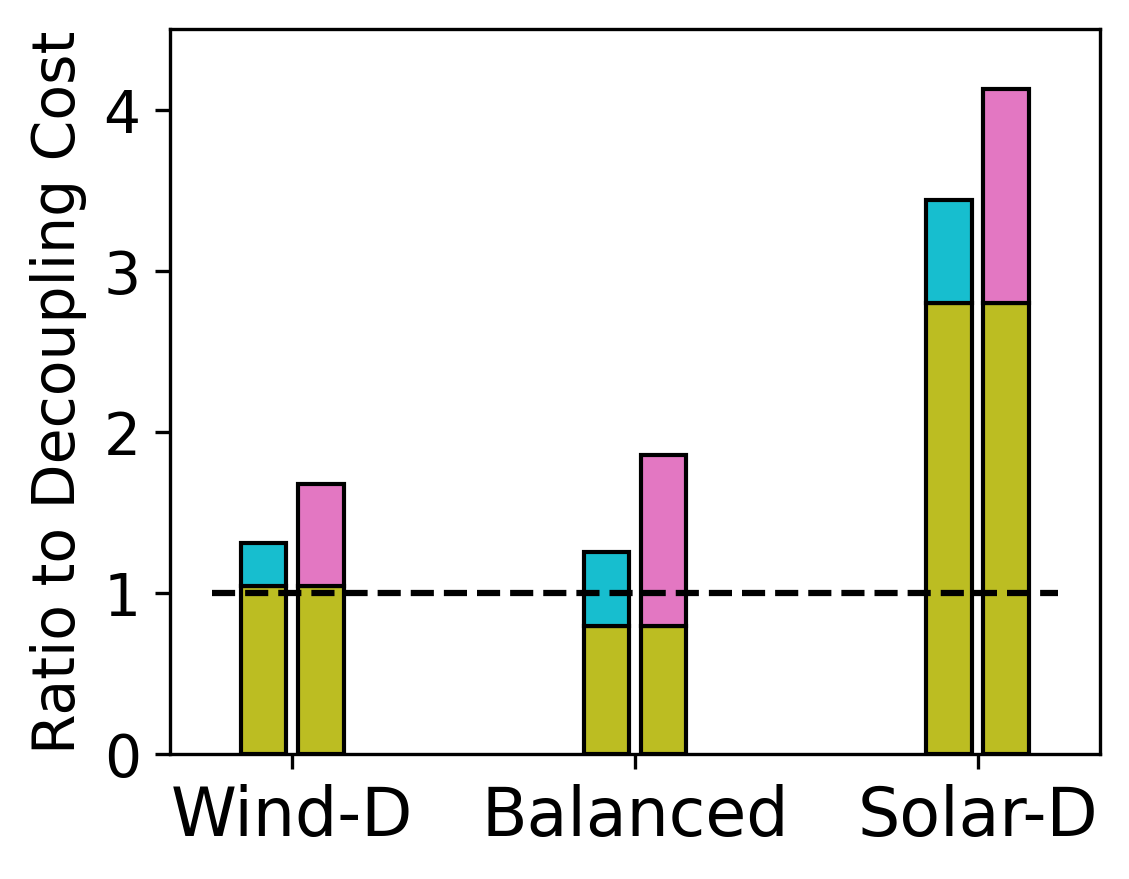}
        \caption{GridCtrl.}
    \end{subfigure}
    \caption{Average Datacenter Benefits vs. Decoupling Cost. (PC-R: power cost reduction, C-R: carbon reduction)}
    \label{fig:DCCostSavingsVsDecoupling}
\end{figure}

With decoupling, datacenters get two economic benefits: (1) reduced power cost and (2) reduced carbon emissions.  We compare the average benefits normalized to average decoupling costs
%benefits the outcomes averaged across datacenters and normalized to 
(70\% total budget, OptDist distribution) in Figure \ref{fig:DCCostSavingsVsDecoupling}. Power cost savings alone are significant. 
With PS-GridScale (Figure \ref{fig:DCCostSavingsVsDecoupling}a, ``PC-R''), the ratio of power cost reduction to decoupling cost is 0.4--3, which is especially high in the solar-dominated grid where the power price is higher. 
%GridCtrl improves the savings to 79--280\% of the decoupling cost. 
With the avoided cost of carbon credits considered, grid metrics-based (GM) allocation brings the total cost savings to 0.8--3.6x of decoupling cost.
%and 125--344\% for GridCtrl. 
Action-based (Act) allocation allocates more grid carbon reduction to datacenters, and directly the average economic benefits to datacenters exceed their decoupling cost. GridCtrl can improve the benefit-cost ratio in the balanced grid by further lowering datacenter power cost.

%To translate carbon reduction into monetary benefit, we assume datacenter operators need to purchase high-quality carbon credit to offset the carbon emissions of datacenters, which begins to be adopted and costs \$280/metric ton CO$_2$ \cite{googleCarbonCapture24}.

\begin{figure}[h]
    \centering
    \begin{subfigure}{.48\columnwidth}
        \centering
        \includegraphics[width=\columnwidth]{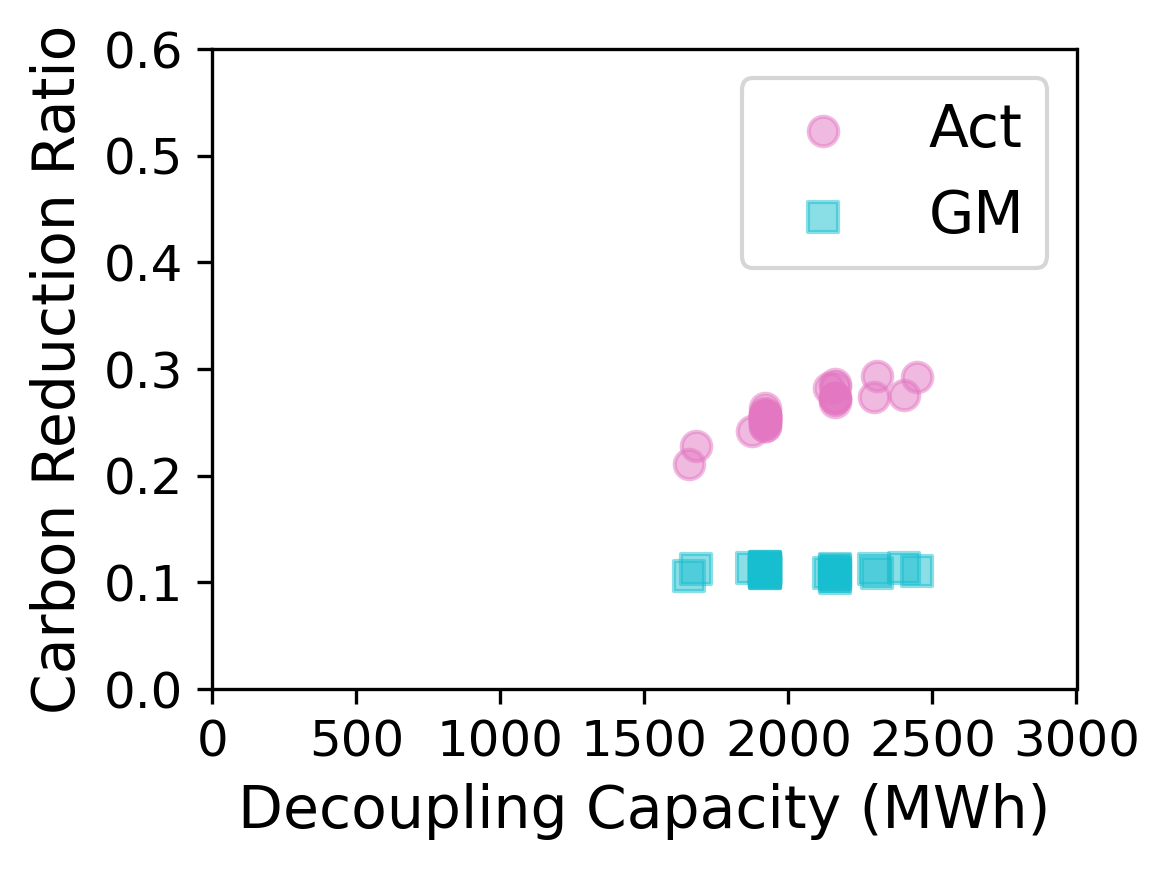}
        \caption{Wind-dominated Grid.}
    \end{subfigure}
    \begin{subfigure}{.48\columnwidth}
        \centering
        \includegraphics[width=\columnwidth]{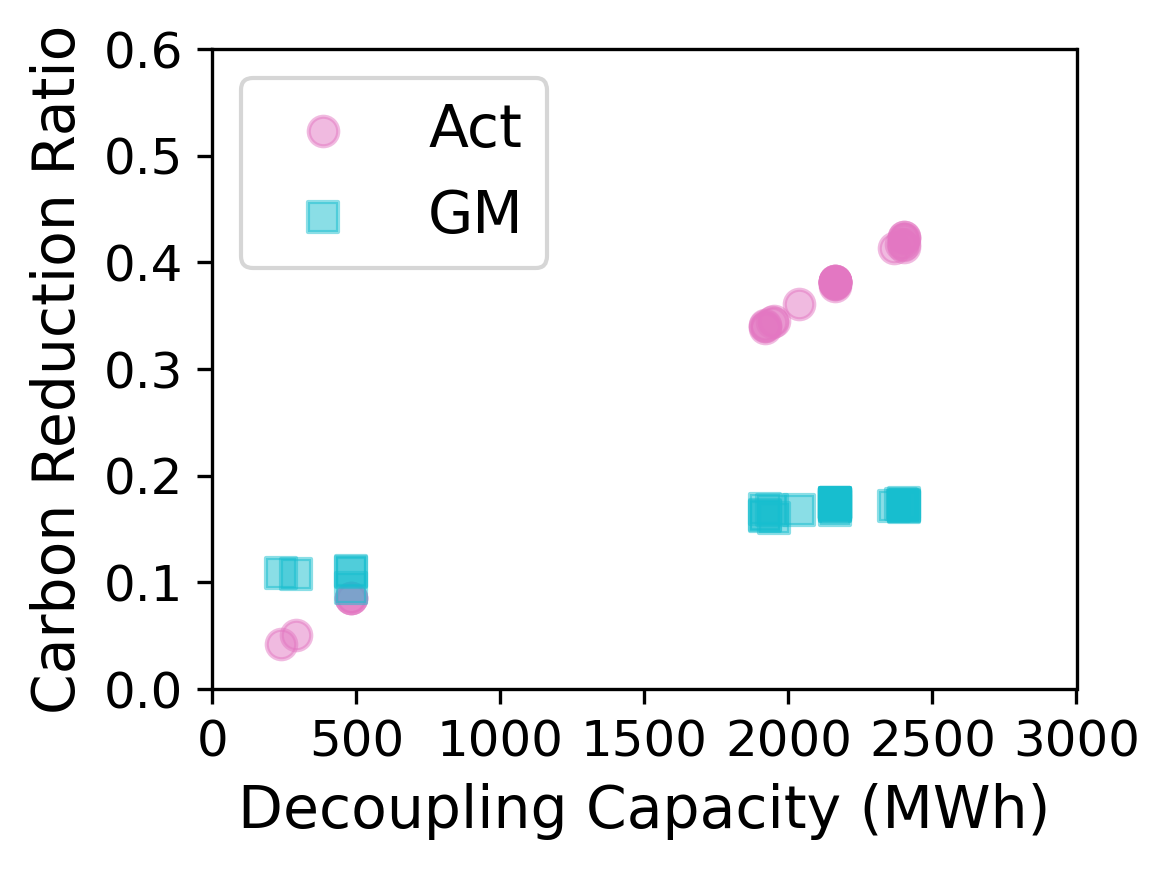}
        \caption{Solar-dominated Grid.}
    \end{subfigure}
    \caption{Carbon Reduction (GridCtrl) with Grid Metrics- and Action-based Allocation at Individual Datacenters.}
    \label{fig:DCCarbonRed}
\end{figure}

We consider individual datacenters' carbon reduction benefits, using the two different reduction allocation methods, GM and Act, in Figure \ref{fig:DCCarbonRed}. With GM allocation, datacenters each see a carbon reduction of about 10\%, mostly due to decreased in grid carbon intensity.  Act allocation attributes all grid carbon reduction to datacenters, proportional to local decoupling capacity.  In this case, most see a >2x reduction, with outliers low due to low local decoupling capacity.

\begin{figure}[h]
    \centering
    \begin{subfigure}{.48\columnwidth}
        \centering
        \includegraphics[width=\columnwidth]{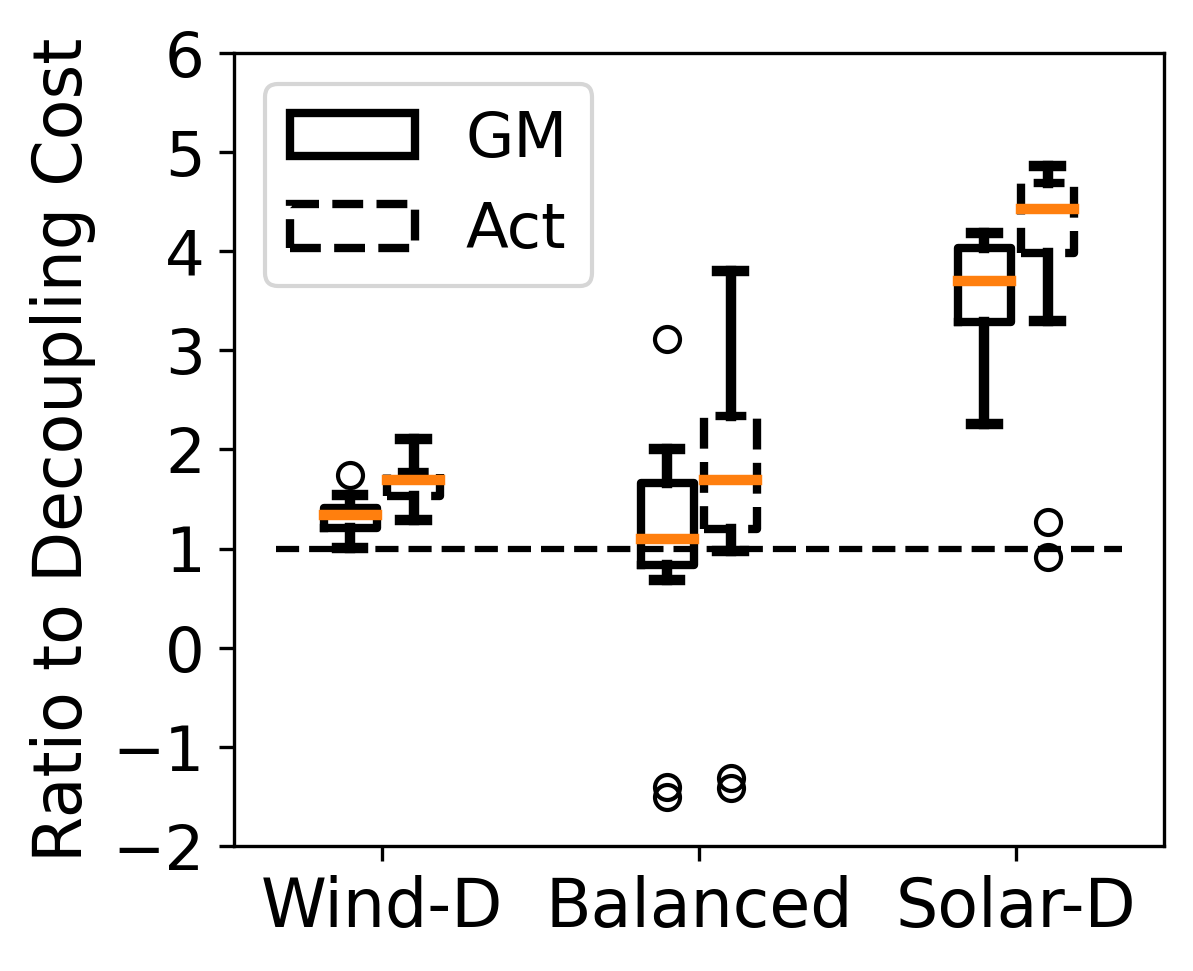}
        \caption{PS-GridScale.}
    \end{subfigure}
    \begin{subfigure}{.48\columnwidth}
        \centering
        \includegraphics[width=\columnwidth]{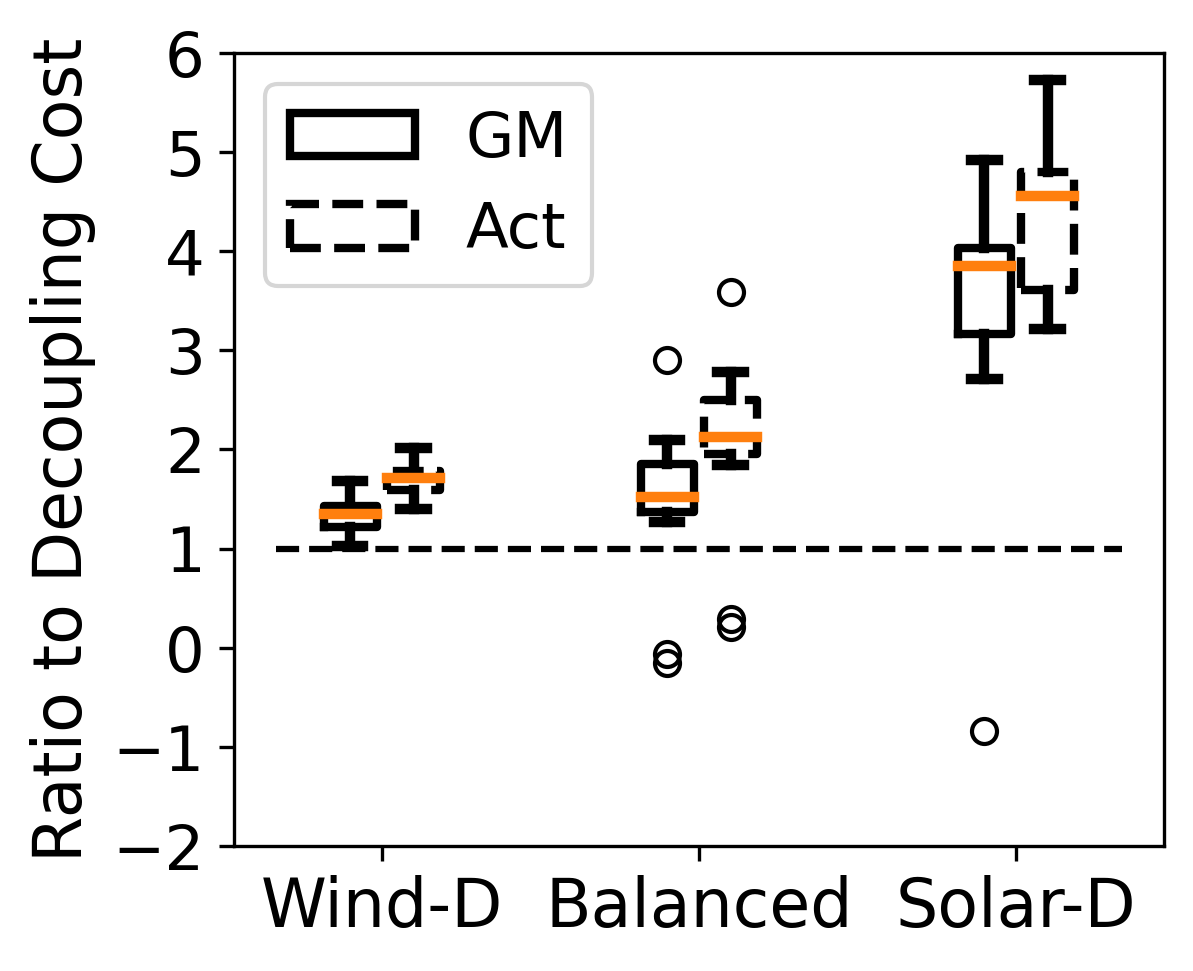}
        \caption{GridCtrl.}
    \end{subfigure}
    \caption{Ratio of Economic Benefits to Decoupling Cost at Individual Datacenters.}
    \label{fig:DCCostSavingsVsDecoupling_DCs}
\end{figure}

Datacenters may have different owners\footnote{And OptDist produces different decoupling investment needs.}, so differences amongst them are important.
Figure \ref{fig:DCCostSavingsVsDecoupling_DCs} compares the ratio of economic benefits (power cost savings and carbon reduction) to decoupling cost at individual datacenters.  In the wind-dominated grid, all datacenters see a net gain (ratio$>$1) from load decoupling with either management approach. In the balanced grid, most datacenters experience an even higher benefit-cost ratio, but there is a wide range.  In the solar-dominated grid, the average ratio is even higher, but there is an even wider range.  In both balanced and solar-dominated grids, some datacenters see very negative returns, and thus are unlikely to deploy decoupling of their own initiative.  Overall, these results suggest that voluntary programs are unlikely to produce optimal (or even good) decoupling distributions, and some differential incentive programs will be needed to achieve OptDist-like distributions.

%In the balanced and solar-dominated grids, while most of the datacenters see improved benefits, there are several outliers experiencing increase in power cost (reflected in the negative ratio). Their distributed decoupling is significantly less than others', implying that the opportunity to exploit low-priced renewables is limited at these locations. The cost increase can be canceled out if the owner also owns datacenters that get normal benefits. Future datacenter site selection may consider this impact. 

\subsubsection{Grid Benefits and Costs}
%Previous evaluation has shown that decoupling can reduce grid carbon emissions, which helps the grid achieve its decarbonization goals \cite{California25MMTCarbonBy2035,eirgrid23report,EUClimateNeutralBy2050,china2023carbonPolicies}. 
An alternative to decoupling is to add more renewable generation, thereby advancing grid decarbonization.  In fact, some studies suggest that overprovisioning renewables is the cheapest way to reduce carbon emissions \cite{MN-solar-curtailment18,gupta2023optimal}.  We compare
the effectiveness of additional renewable generation against decoupling.

%Grid renewable capacity is added following the wind-solar ratio in different types of grids. 

\begin{figure}[h]
    \centering
    \begin{subfigure}{.48\columnwidth}
        \centering
        \includegraphics[width=\columnwidth]{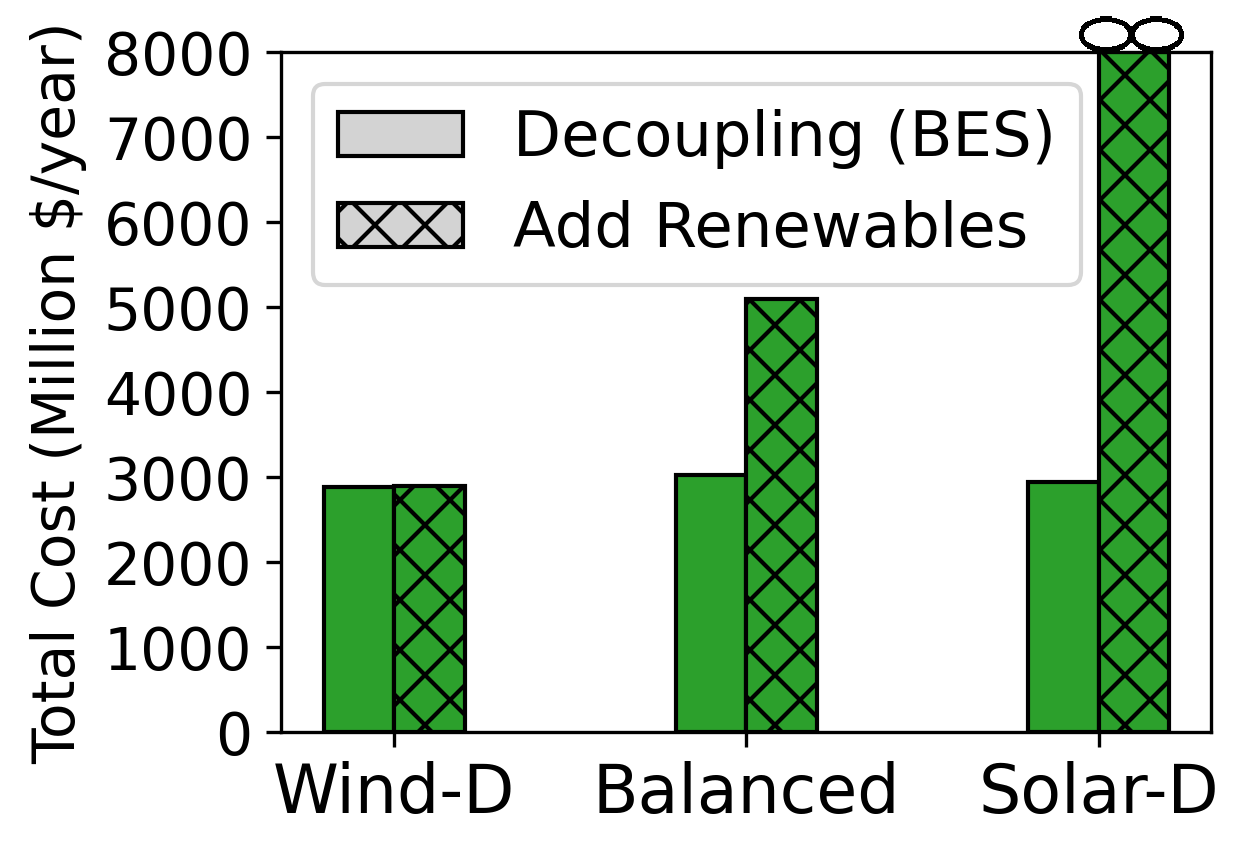}
        \caption{PS-GridScale.}
    \end{subfigure}
    \begin{subfigure}{.48\columnwidth}
        \centering
        \includegraphics[width=\columnwidth]{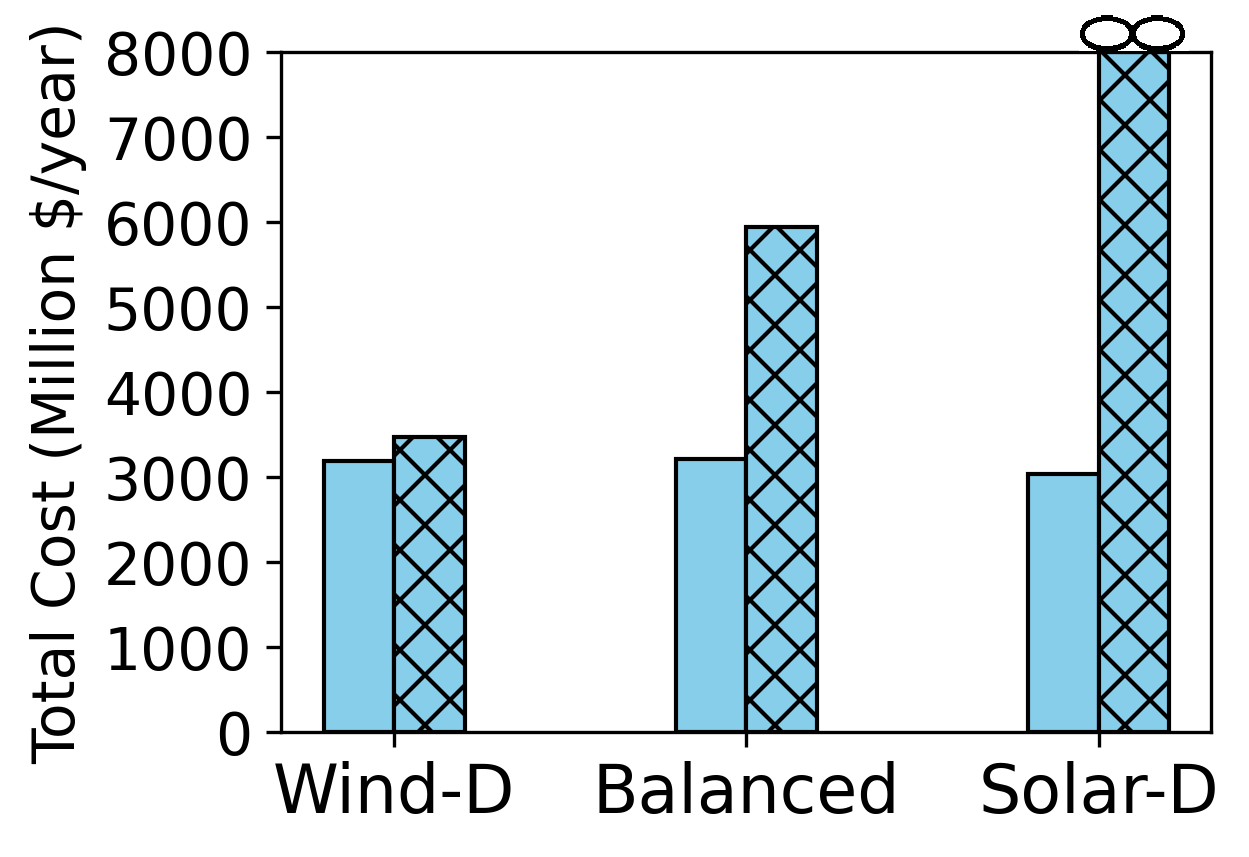}
        \caption{GridCtrl.}
    \end{subfigure}
    \caption{Comparison of Grid Carbon Reduction Cost ($\infty$: carbon reduction unachievable with adding renewables).}
    \label{fig:carbonRedCostVsRenewGen}
\end{figure}

Figure \ref{fig:carbonRedCostVsRenewGen} compares the annualized generation cost\footnote{Estimated with levelized cost of energy (LCOE) of \$50/MWh for wind and \$60/MWh for solar (2023 midpoints \cite{lazard2025LCOE}).} with decoupling (70\% total budget, OptDist distribution). With PS-GridScale, 
the costs are the same in wind-dominated grid, but decoupling is 40\% cheaper in balanced grid and enables reduction unachievable with only more solar generation. GridCtrl slightly increases the decoupling cost due to more battery usage, but as the carbon benefits improve more, its decoupling cost is 8\% lower in the wind-dominated grid and 46\% lower in the balanced grid. 

Overall, for grids with 60\% renewables, implementing datacenter load decoupling (flexibility) is more economic for decarbonization than adding more renewables. 

\section{Related Work}
\label{sec:related}

\subsection{Datacenter Flexibility}
\paragraph{Load Decoupling} Previous work that controls energy resources co-located with datacenters for load adaptation shares the idea of load decoupling. Some explore reusing widely-equipped UPS energy storage for grid services such as frequency regulation and peak shaving, reducing power cost and aiding the grid \cite{shi2016leveraging,urgaonkar2011optimal,zhang2021distributed,msGridUPS}, but the energy capacity is too low (typically $\approx$10 minutes) to decouple datacenter power and grid load for hours---scenarios in this paper. \cite{ren2012carbon,acun2023carbon} assessed how much renewable power purchase agreement (PPA) and energy storage would be needed for various renewable matching goals (e.g. 100\% hourly matching) at individual DCs.  We do extensive  DC-grid coupled simulations that capture complex dynamics to explore decoupling needs and examine both DC and grid benefits.
%achieving significant grid benefits by exploring distribution and management of decoupling.

\paragraph{Workload Flexibility} Although we assume datacenter power capacity needs to be constant as is typical with cloud computing and important for high DC capital efficiency, some workloads are  time- and space-shiftable (e.g. machine learning model training and batch processing jobs).  Shifting enables flexible DC capacity without QoS loss.  These workloads can be significant in some 
%Such workloads account for a large fraction of 
commercial cloud workloads (e.g. 25\% of CPU core usage at Microsoft) \cite{acun2023carbon,parayil2024towards,radovanovic2021carbon}. %with growth %and may continue growing with the need of LLM model training. 
Many researchers have exploited scheduling or scaling these workloads to match datacenter grid load to low-priced/low-carbon generation for cost or carbon reduction \cite{sukprasert2024limitations,wiesner2021let,luo2013temporal,goiri2013parasol,acun2023carbon,radovanovic2021carbon,hanafy2023carbonscaler,lin2012dynamic,liu2012renewable,xing2023carbon}.  However, all of these studies assume space capacity, an undesirable situation from an economic or business point-of-view.
Workload flexibility is complementary to load decoupling, and correct exploitation can reduce both energy deficit and decoupling cost.

\subsection{Coordinating Adaptive Datacenters and Grid}
Our exploration of managing decoupling for efficiency is related to coordinating DC flexibility and grid dispatch. \cite{lin2024exploding,norris2025rethinking,ShedShiftLBNL24} show the benefits of exploiting load flexibility for grid resource adequacy or datacenter growth but do not address the coordination problem. \cite{lin2021evaluating,lindberg2021guide,gorka2025electricityemissions} show that selfish control (local optimization) can harm the grid and datacenters themselves when DCs consume 10--20\% of grid energy (or more).  With the growth of ML/AI, more power grids are seeing DC load fractions that exceed 15\% \cite{ercotLoadForecast,epriDCLoadForecast2024} (or even 30\% \cite{dominionVA23report,eirgrid23report}), and thus coordination is becoming increasingly important. %GridCtrl can be viewed as an extension to demand response \cite{liu2013data,le2016joint,zhou2018truthful}, where DCs adjust load following grid signals, with regular grid control and large load variation up to $\pm 30\%$.
Several studies of coordination explore information sharing and 
various DC-grid
interaction designs, studying load adaptation with grid simulation \cite{lindberg2021guide,zhang2020flexibility,menati2023modeling,wu2023incentivizing,lin2023adapting,gorka2025electricityemissions,zhou2018truthful}. Specifically, we show that 1-way info sharing represented by PlanShare \cite{lin2023adapting} can be ineffective as DCs continue to grow.  Addressing this, we design PS-GridScale that empowers the grid to participate in load shaping, enabling it to outperform PlanShare significantly.
%by introducing reactive grid control.

\section{Summary and Future Work}
\label{sec:summary}
With the rapid growth of datacenter load, DC load flexibility is critical in reducing datacenter carbon emissions.  Specifically, it can resolve the conflict between DC growth and grid decarbonization.  
To protect DC compute efficiency and QoS, such flexibility requires decoupling compute capacity and grid load.  With the goal of achieving high benefit at low cost, we study how to best distribute and manage decoupling. Our evaluation shows that >98\% of maximum benefits can be achieved with 70\% of the maximum decoupling need.  Cooperation between the grid and DCs is necessary for efficient decoupling management: 2-way sharing and control represented by PS-GridScale enables 1.4x grid carbon reduction vs. 1-way information sharing and achieves 84--90\% of the optimal benefits.  Finally, we show that decoupling can be economically viable, the economic benefits to datacenters, on average, exceed their decoupling costs.
However, significant skew across different datacenters---some may not benefit at all or enough to defray their costs---means a significant grid intervention may be required to implement decoupling.

Our framework of datacenter load decoupling opens up many research opportunities. When datacenters propose load profiles, can there be better grid metrics that guide effective shaping and avoid oversubscribing the opportunities? Allocation methods that fairly allocate the grid benefits and are invulnerable to malicious load profiles are important for formulating grid programs. Discussion beyond technical problems such as who should pay for decoupling is also important.
%Finally, it may be time to revisit the datacenter site selection and grid resource planning studies.

%\subsubsection{Summary and Discussion}
%We evaluated the benefits and costs directly resulting from load decoupling, showing that both the grid and datacenters have economic incentives to implement decoupling. However, there is consideration beyond this comparison. Who has the responsibility and more power to implement decoupling? Is it fair to let the grid pay for the solution to problems caused by datacenters? For hyperscalers, the protection of quality of service from flexed grid load and ability to build datacenters faster are (probably more) important---the business with datacenters is so profitable that they already begin to add energy resources without the economic benefits evaluated but for regulatory requirements \cite{msftIEGasGenforDC,xAIMegapack25}.

\begin{acks}
This work is supported in part by NSF Grants OAC-2442555, CNS-2325956, and the VMware University Research Fund. We also thank the Large-scale Sustainable Systems Group members for their support of this work!
\end{acks}

\clearpage
\bibliographystyle{ACM-Reference-Format}
\bibliography{Bib/dcSim, Bib/zccloud-10-2020, Bib/middlebox, Bib/dissertation}

\appendix
\section{Formulation and Settings of Direct-current Optimal Power Flow (DC-OPF)}
\label{appendix:gridSimDetails}
Here we present the complete formulation of DC-OPF problem used for grid simulation, starting from the notations:

\begin{table}[h]
\caption{DC-OPF Notations: Sets}
\label{tab:notation-set}
\begin{center}
\scalebox{0.85}{
\begin{tabular}{p{35pt}|p{85pt}|p{35pt}|p{85pt}}
  \hline
  Notation & Description & Notation & Description\\
  \hline
  $\mathcal{T}$ & Time periods & $\mathcal{N}$ & Buses\\ 
  $\mathcal{G}$ ($\mathcal{G}_n$) & Generators (at bus $n$) & $\mathcal{I}$ ($\mathcal{I}_n$) & Import points (at bus $n$)\\
  $\mathcal{L}$ & Transmission lines & $\mathcal{L}_n^+/\mathcal{L}_n^-$ & Transmission lines to/from bus $n$\\
  $\mathcal{DC}$ ($\mathcal{DC}_n$) & Datacenters (at bus $n$) & $\mathcal{ND}$ ($\mathcal{ND}_n$) & Non-DC loads (at bus $n$)\\
  $\mathcal{W}$ ($\mathcal{W}_n$) & Wind farms (at bus $n$) & $\mathcal{S}$ ($\mathcal{S}_n$) & Solar farms (at bus $n$)\\
  $\mathcal{R}$ ($\mathcal{R}_n$) & Other renewable generators (at bus $n$) & & \\
  \hline
\end{tabular}
}
\end{center}
\end{table}

\begin{table}[h]
\caption{DC-OPF Notations: Decision Variables}
\label{tab:notation-dv}
\begin{center}
\scalebox{0.85}{
\begin{tabular}{p{40pt}|p{80pt}|p{40pt}|p{80pt}}
  \hline
  Notation & Description & Notation & Description\\
  \hline
  $p_{i,t}$ & Generation of generator $i$ at time $t$ & $f_{l,t}$ & Power flow of line $l$ at time $t$\\
  $d_{i,t}^{nd}$ & Load shedding at non-DC load $i$ at time $t$ & $d_{i,t}^{dc}$ & Load shedding at datacenter $i$ at time $t$\\
  $m_{i,t}$ & Curtailment at import $i$ at time $t$ & $w_{i,t}$ & Curtailment at wind farm $i$ at time $t$\\
  $s_{i,t}$ & Curtailment at solar farm $i$ at time $t$ & $r_{i,t}$ & Curtailment at other renewable $i$ at time $t$\\
  $\theta_{n,t}$ & Phase angle at bus $n$ at time $t$ &  & \\
  \hline
\end{tabular}
}
\end{center}
\end{table}

\begin{table}[h]
\caption{DC-OPF Notations: Parameters}
\label{tab:notation-param}
\begin{center}
\scalebox{0.85}{
\begin{tabular}{p{35pt}|p{85pt}|p{35pt}|p{85pt}}
  \hline
  Notation & Description & Notation & Description\\
  \hline
  $B_l$ & Susceptance of transmission line $l$ & $C_i$ & Generation cost of generator $i$\\
  $C_i^{nd}$ & Load-shedding penalty at non-DC load $i$ & $C_i^{dc}$ & Load-shedding penalty at datacenter $i$\\
  $C_i^w$ & Curtailment penalty at wind farm $i$ & $C_i^s$ & Curtailment penalty at solar farm $i$\\
  $C_i^m$ & Curtailment penalty at import point $i$ & $C_i^r$ & Curtailment penalty at other renewable $i$\\
  $D_{i,t}$ & Power demand of load $i$ at time $t$ & $F^{max}_l$ & Maximum power flow of transmission line $l$\\
  $M_{i,t}$ & Power from import $i$ at time $t$ & $P^{max}_i$ & Maximum power output of generator $i$\\
  $RD_i$ & Ramp-down limit of generator $i$ & $RU_i$ & Ramp-up limit of generator $i$\\
  $W_{i,t}$ & Generation of wind farm $i$ at time $t$ & $S_{i,t}$ & Generation of solar farm $i$ at time $t$\\
  $R_{i,t}$ & Generation of other renewable $i$ at time $t$ & $\Theta_{n,t}^{min}$ & Minimum phase angle at bus $n$ at time $t$\\ $\Theta_{n,t}^{max}$ & Maximum phase angle at bus $n$ at time $t$ & & \\
  \hline
\end{tabular}
}
\end{center}
\end{table}

The generation cost of thermal generators is 6/31/22 \$/MWh for nuclear/coal/gas generators \cite{eiaFuelCost}. Penalties of load shedding and curtailment are listed in Table \ref{tab:paramSettings}, which are reflected in the locational marginal price (LMP) when load shedding or renewable/import curtailment happens. The other parameters are set according to load/generation profiles or static infrastructure attributes.

\begin{table}[h]
\caption{DC-OPF Parameter Settings}
\label{tab:paramSettings}
\begin{center}
\scalebox{0.85}{
\begin{tabular}{p{20pt}|p{45pt}|p{20pt}|p{45pt}|p{20pt}|p{45pt}}
  \hline
   & Value &  & Value &  & Value\\
  \hline
  $C_i^{nd}$ & \$1000/MWh & $C_i^{dc}$ & \$1000/MWh & $C_i^w$ & \$100/MWh\\
  $C_i^s$ & \$100/MWh & $C_i^m$ & \$500/MWh & $C_i^r$ & \$1000/MWh\\
  \hline
\end{tabular}
}
\end{center}
\end{table}

The optimization objective is to minimize the daily dispatch cost consisting of generation cost, load shedding penalties, and import/renewable generation curtailment penalties:
\begin{subequations}
\label{eq:ED}
\begin{align}
  \boldsymbol{\min} \  
  & \sum_{t\in\mathcal{T}} \left( \sum_{i\in\mathcal{G}} C_i p_{i,t} + \sum_{i\in\mathcal{ND}} C_i^{nd} d_{i,t}^{nd} + \sum_{i\in\mathcal{DC}} C_i^{dc} d_{i,t}^{dc} + \sum_{i\in\mathcal{I}} C_i^m m_{i,t} \right. \notag \\
  & \qquad \left. + \sum_{i\in\mathcal{W}} C_i^w w_{i,t} + \sum_{i\in\mathcal{S}} C_i^s s_{i,t} + \sum_{i\in\mathcal{R}} C_i^r r_{i,t} \right) \label{eq:ED:obj}
\end{align}
subject to typical constraints including balancing at each node (\ref{eq:ED:flow}), transmission (\ref{eq:ED:angle}--\ref{eq:ED:anglerange}), generator capacity (\ref{eq:ED:maxgen}) and ramping (\ref{eq:ED:ramp}), and shedding/curtailment limits (\ref{eq:ED:maxshed}--\ref{eq:ED:maxrenewable}):
\begin{align}
  \text{s.t.} \quad & \sum_{l\in\mathcal{L}_n^+} f_{l,t} - \sum_{l\in\mathcal{L}_n^-} f_{l,t} + \sum_{i\in\mathcal{G}_n} p_{i,t} + \sum_{i\in\mathcal{I}_n} (M_{i,t} - m_{i,t}) \notag \\
  & + \sum_{i\in\mathcal{W}_n} (W_{i,t} - w_{i,t}) + \sum_{i\in\mathcal{S}_n} (S_{i,t} - s_{i,t}) + \sum_{i\in\mathcal{R}_n} (R_{i,t} - r_{i,t}) \notag \\
  & = \sum_{i\in\mathcal{ND}_n} (D_{i,t} - d_{i,t}^{nd}) + \sum_{i\in \mathcal{DC}_n} (gridLoad_{i,t} - d_{i,t}^{dc}), \notag \\ 
  & \quad \forall n\in\mathcal{N}, t\in\mathcal{T}, \label{eq:ED:flow} \\
  & f_{l,t} = B_l (\theta_{n,t} - \theta_{m,t}), \quad \forall l=(m,n)\in\mathcal{L}, t\in\mathcal{T}, \label{eq:ED:angle}\\ 
  & -F_l^{max} \leq f_{l,t} \leq F_l^{max}, \quad \forall l\in\mathcal{L}, t\in\mathcal{T}, \label{eq:ED:linecap}\\
  & \Theta_n^{min} \leq \theta_{n,t} \leq \Theta_n^{max} \quad \forall n\in\mathcal{N}, t\in\mathcal{T}, \label{eq:ED:anglerange} \\
  & 0 \leq p_{i,t} \leq P_i^{max}, \quad \forall i\in\mathcal{G}, t\in\mathcal{T}, \label{eq:ED:maxgen}\\
  & -RD_i \leq p_{i,t} - p_{i,t-1} \leq RU_i, \quad \forall i\in\mathcal{G}, t\in\mathcal{T}, \label{eq:ED:ramp} \\
  & 0 \leq d_{i,t}^{nd} \leq D_{i,t}, \quad \forall i\in\mathcal{ND}, t\in\mathcal{T}, \label{eq:ED:maxshed} \\
  & 0 \leq d_{i,t}^{dc} \leq gridLoad_{i,t}, \quad \forall i\in\mathcal{DC}, t\in\mathcal{T}, \label{eq:ED:maxdcshed} \\
  & 0 \leq m_{i,t} \leq M_{j,t}, \quad \forall i\in\mathcal{I}, t\in\mathcal{T}, \label{eq:ED:maximport} \\
  & 0 \leq w_{i,t} \leq W_{j,t}, \quad \forall i\in\mathcal{W}, t\in\mathcal{T}, \label{eq:ED:maxwind} \\
  & 0 \leq s_{i,t} \leq S_{j,t}, \quad \forall i\in\mathcal{S}, t\in\mathcal{T}, \label{eq:ED:maxsolar} \\
  & 0 \leq r_{i,t} \leq R_{j,t}, \quad \forall i\in\mathcal{R}, t\in\mathcal{T}. \label{eq:ED:maxrenewable}
\end{align}
\end{subequations}

\section{Total Cost of Ownership (TCO) of Battery Energy Storage}
\label{appendix:BESTCO}
Total cost of ownership (TCO) includes capital expenses (CapEx) depreciation and operational expenses (OpEx) incurred periodically. The CapEx of battery storage can be broken down into energy component (EC) and power component (PC) costs, which correspond to batteries and electrical infrastructure, with costs proportional to the energy and power capacity respectively. 

The OpEx of Li-ion batteries covers operation, maintenance, and capacity augmentation to mitigate battery degradation. We assume linear battery degradation by cycle, under which one cycle is counted when accumulated discharged energy reaches $energyCap*DoD$ (depth of discharge). Based on all above, the TCO of battery energy storage (\$/year) can be calculated by:
\begin{equation*}
    \begin{split}
        &\frac{power_{max}*unitCapex_{PC}+energyCap*unitCapex_{EC}}{DeprPeriod_{MB}}\\
        +&power_{max}*unitCapex_{sys}(0.01*cycle_{avg}+0.015)+lossCost
    \end{split}
    \label{eq:BESTCO}
\end{equation*}
where $unitCapex_{sys}=unitCapex_{PC}+unitCapex_{EC}\cdot duration$ is a system cost metric. $lossCost$ is calculated based on typical round-trip efficiency, daily discharge, and power price. Table \ref{tab:tco_BES_param} lists the settings used in Section \ref{sec:evaluation}.

\begin{table}[H]
\caption{Li-ion BES Cost Settings (Sources: \cite{cole2023cost, newell2022pjm})}
\label{tab:tco_BES_param}
\begin{center}
\begin{tabular}{p{86pt}|p{134pt}}
  \hline
  %Attribute & Value(s)\\
  %\hline
  \multicolumn{2}{l}{Capex (80\% DoD, 85\% RTE, $DeprPeriod_{MB}$=15 years)}\\
  \hline
  Power Components & 0.36 million \$/MW (2023)\\
  Energy Components & 0.39 million \$/MWh (2023)\\
  \hline
  \multicolumn{2}{l}{Opex (million \$/(MW$\cdot$year))}\\
  \hline
  Capacity Augmentation & 1\%$\cdot unitCapex_{sys}$$\cdot$\#cycles/day\\
  Others & 1.5\%$\cdot unitCapex_{sys}$\\
  \hline
\end{tabular}
\end{center}
\end{table}

\end{document}